%% file: main.tex
\documentclass[runningheads]{llncs}
\usepackage[T1]{fontenc}
\usepackage{graphicx}
\usepackage{hyperref}
\usepackage{color}

\usepackage{hhline}
\usepackage{cleveref}
\usepackage{scrextend}
\usepackage[labelformat=simple]{subcaption}
\usepackage{footnote}
\makesavenoteenv{table}
\makesavenoteenv{tabular}
\usepackage{makecell}

\usepackage{listings}
\usepackage{xcolor}
\usepackage{comment}

\definecolor{codegreen}{rgb}{0,0.6,0}
\definecolor{codegray}{rgb}{0.5,0.5,0.5}
\definecolor{codepurple}{rgb}{0.58,0,0.82}
\definecolor{backcolour}{rgb}{0.95,0.95,0.92}

\lstdefinestyle{mystyle}{
  backgroundcolor=\color{backcolour}, commentstyle=\color{codegreen},
  keywordstyle=\color{magenta},
  numberstyle=\tiny\color{codegray},
  stringstyle=\color{codepurple},
  basicstyle=\ttfamily\footnotesize,
  breakatwhitespace=false,         
  breaklines=true,                 
  captionpos=b,                    
  keepspaces=true,                 
  numbers=left,                    
  numbersep=5pt,                  
  showspaces=false,                
  showstringspaces=false,
  showtabs=false,                  
  tabsize=2
}
\input{solidity-highlighting.tex}

\usepackage{authblk}
\usepackage{biblatex}
\begin{document}
%

\title{The Hidden Shortcomings of (D)AOs --\\ An Empirical Study of On-Chain Governance}
\titlerunning{The Hidden Shortcomings of (D)AOs}
%
\author{Rainer Feichtinger \and Robin Fritsch \and Yann Vonlanthen \and Roger Wattenhofer}
\institute{ETH Zurich \\ \email{\{rainerfe,rfritsch,yvonlanthen,wattenhofer\}@ethz.ch}}
%
\authorrunning{Feichtinger et al.}
%
%
\maketitle              
\begin{abstract}
Decentralized autonomous organizations (DAOs) are a recent innovation in organizational structures, which are already widely used in the blockchain ecosystem.
We empirically study the on-chain governance systems of 21 DAOs and open source the live dataset. The DAOs we study are of various size and activity, and govern a wide range of protocols and services, such as decentralized exchanges, lending protocols, infrastructure projects and common goods funding. 
Our analysis unveils a high concentration of voting rights, a significant hidden monetary costs of on-chain governance systems, as well as a remarkably high amount of pointless governance activity.

\end{abstract}

\section{Introduction}

Decentralized autonomous organizations (DAOs) are already an integral part of today's blockchain ecosystem, and, as a significant innovation in organizational structures, they have the potential of impacting vast parts of business and society in the future \cite{wef2022daos}.
In early 2023, DAOs are estimated to manage an equivalent of 12 Billion US dollars\footnote{\url{https://deepdao.io/organizations}}.
Their goal is to enable open deliberation in a decentralized community, allow for transparent voting, and democratize resources.
This has the potential of resulting in fast and well accepted decisions, something that today's centralized decision-making lacks.

With the first DAOs emerging very recently and most of them being less than a few years old, DAOs are still very much re-inventing themselves. In this work, we aim to shed light on some areas in which DAOs are facing challenges today. We take a data-first approach, and look at 21 DAOs with on-chain governance systems in depth, making the collected dataset openly accessible for each.
We observe large scale phenomena, such as low decentralization of voting power, and uncover more overlooked topics, such as the high cost of DAOs performing votes on a popular blockchain such as Ethereum.
Paired with a very concentrated token distribution, the question of the purpose of the DAO beyond a pure marketing tool might have to be reconsidered.

In this paper, we investigate DAOs that have deployed a governance system on a blockchain to reach collective decision.
These DAOs issue a governance token which represents voting rights, and distribute the token among stakeholders and the community (possibly via an airdrop or a token sale).
Holders of the governance token can decide to delegate their voting rights to any representative (including themselves) to represent their interests.
These so-called delegates can then use this voting power to vote on proposals which are brought forward.

The contribution of this paper is twofold:
First, we make an extensive dataset on the governance systems of 21 DAOs easily accessible. The data contains a complete history of all token holders, delegations, proposals and votes.
Second, we analyze the acquired dataset to get a comprehensive overview of the state of on-chain governance systems.

The first point might seem surprising, since one of the most appealing promises of blockchains is transparency: all transactions are public and can be viewed by anyone at any time.
However, in practice it is not trivial to acquire all governance related information from raw blockchain data. 
To make the data accessible, we create a so-called \emph{subgraph} for each of the DAOs using The Graph protocol\footnote{\url{https://thegraph.com/en/}}. These subgraphs allow retrieving all governance-related data though a GraphQL API.
Since each governance system has its specificities, data collection had to be individually adapted to the particularities of each smart contract.
All subgraph code is open source to allow for extending the dataset to include more DAOs by adding further subgraphs \cite{feichtinger23git}.

When analyzing the data, we focus on the distribution of voting rights and the monetary cost of the governance systems, among other aspects.
While we do see a slight trend towards decentralization, voting power is still highly centralized in most DAOs.
Indeed, for 17 out of the 21 analyzed governance systems, a majority of voting power, which suffices to decide any vote, is controlled by less than 10 participants.
Furthermore, in most DAOs most voting power is held by delegates mainly representing a single token holder. Hence, there is little evidence of a substantial community-participation in the decision-making.
Moreover, we quantify the monetary cost of the governance systems, in terms of transaction costs for delegating and voting, but also in added overhead for token transfers. This unveils significant costs, up to millions of dollars for some DAOs.
Surprisingly, we also discover numerous pointless transactions, and their frequency even increasing in some cases -- a sign of an immature governance systems.

\section{Related Work}

The history of DAOs on blockchains goes back to 2016 when a first DAO (called ``The DAO'') was formed on Ethereum.
Unfortunately however, before becoming operational, the project suffered a severe hack which drained deposited funds from the DAO \cite{dupont2017thedao}. (The event was so significant that it lead to a hard fork of the Ethereum blockchain and the creation of Ethereum Classic).

After this failure, it took a few years before the idea of DAOs gained traction again.
In 2018, the stablecoin protocol MakerDAO introduced an on-chain governance system as one of the first blockchain-based applications \cite{makerdao2018governance, makerdao2015mkr} (see \cite{sun2022makerdao} for an empirical study).

The next wave of DAOs then started entering the stage in 2020, kicked off by Compound finance \cite{leshner2020governance}.
Since then, more and more blockchain-based applications have followed suit and introduced an on-chain governance system, among them many decentralized finance (DeFi) protocols. (An overview and more background on DeFi can be found in \cite{zetsche20defi, werner2021defi_sok, aramonte2021defi}.)

On the empirical side, there are already a number of studies of these governance systems.
Some brief and early ones, such as \cite{stroponiati2020, nadler2020defi_ownership, jensen2021decentralized, barbereau2022distribution}, focus mainly the distribution of the ownership of governance tokens.
In a more detailed study, Barbereau et al.\ consider the governance systems of nine DeFi protocols including MakerDAO, Compound and Uniswap \cite{barbereau2022decentralised}. Besides the token distribution, they also examine the voter turnout on governance decisions.
More recently, a more comprehensive analysis additionally examined the voting behavior and the structure of the delegation network for three governance systems: Compound, Uniswap and ENS \cite{fritsch2022daos}.

In contrast to this line of empirical work, Aoyogi and Ito \cite{aoyagi22competing_daos} define a theoretical model of DAOs and study the competition of platforms with decentralized and centralized governance.
DAOs have also been examined from a regulatory and compliance angle such as in \cite{axelsen2022daos}, or in \cite{ding2022DAOgovernance} which includes a case study of GnosisDAO.
A qualitative comparison of DAO platforms can be found in \cite{baninemeh2023daos}.

The topic is also reaching mainstream attention with the World Economic Forum publishing two extensive reports on DAOs including parts on their strengths and weaknesses, the keys risks, operational processes, DAO governance processes as well as major legal and regulatory questions DAOs must face \cite{wef2022daos, wef2023daos}. 

The voting systems used by most DAOs apply elements of \emph{liquid democracy} (sometimes also referred to as \emph{delegative democracy}) \cite{ford2002delegative, behrens2017liquid, zuber17liquid, valsangiacomo22liquid}, in particular the possibility of delegating voting power to delegates.
The main difference to liquid democracy is that most DAOs use the plutocratic ``one token, one vote'' approach instead of following the classic ``one person, one vote'' principle as outlined in liquid democracy.
A second difference is that delegating is not transitive as it is in most forms of liquid democracy. (Delegations being transitive means that delegates can again delegate voting power they received by delegation to another delegate, and so on.)
Instead, the systems currently implemented by DAOs only permit a single delegation step from a token holder to a delegate.

The most prominent use case of liquid democracy is an internal voting system used by the German Pirate Party which has been studied in \cite{swierczek2011liquid, kling2015voting}.

There are also parallels between DAO governance's ``one token, one vote'' and the ``one share, one vote'' principle of shareholder democracy \cite{mitchell2006shareholder}.
In this sense, DAO governance decisions (specially for DeFi protocols) can be seen as an equivalent to decisions at shareholder meetings of traditional companies. 
The voting behavior in traditional shareholder meeting has been studied in the literature, e.g.\ \cite{li2019shareholders, zachariadis2020free}.

\section{Methodology and Dataset}

We analyze a total of 21 on-chain governance systems that run atop the Ethereum blockchain, including systems that govern decentralized exchanges (Uniswap), lending protocols (Compound, Silo, Inverse, Euler), infrastructure (ENS, Radicle), services (GasDAO, Instadapp, Braintrust), and public goods funding (Gitcoin). A more detailed overview of the DAOs we analyzed can be found in Appendix \ref{appendix:daos}.

\subsection{Data Collection}

We collect data by using the open source platform called \emph{The Graph}. The data for each individual DAO is indexed with a so-called subgraph. As each protocol has its specificities, we tailored each subgraph accordingly. Once indexed, these subgraphs allow the retrieval of the pre-defined data through a GraphQL API. All our subgraphs have been open-sourced \cite{feichtinger23git}
\footnote{The live subgraphs can also be found at \url{https://thegraph.com/hosted-service/subgraph/governancedao/NAME-governance} where NAME is to be replaced by the name of the DAO.}.

For each DAO, the subgraphs store data on all token holders (their address, their token balance, and the address they are delegating to) and all delegates (their address, the amount of votes delegated to them, and which holders delegate to them).
All this information is retrievable at an arbitrary block height.
Furthermore, the subgraphs contain details on all delegation transactions (who delegated when to whom) and on all votes cast by delegates (how, when, on which proposal, and with now much voting power the delegate voted).
All transfers of the governance tokens are also stored, since they are necessary to determine how much voting power a delegate holds at any given time.
Finally, metadata on the governance systems and all proposals is included.

\subsection{Dataset}

Before further analyzing the data, we filter the dataset by removing smart contract accounts and accounts managed by an exchange. Usually, tokens in these addresses are not controlled by any single user, and they cannot be used in governance.
For example, a single smart contract\footnote{address: \textit{0xd7a029db2585553978190db5e85ec724aa4df23f}} holds about 61\% of the ENS tokens at the time of writing. This contract is a time lock for ENS tokens that will only be available over the next few years. Therefore, no one can currently participate in governance with these tokens. 
Including such accounts in the analysis would strongly distort the results, especially when analyzing the distribution of voting power.
We queried whether an account is a smart contract or not via Alchemy\footnote{https://alchemy.com} and we retrieved a list of accounts controlled by an exchange from Etherscan\footnote{https://etherscan.io/accounts/label/exchange}.

Table \ref{table:data_overview} shows an overview of the analyzed DAOs. For each, we consider the time period between its deployment and block 16,530,000 (31 Jan 2023).

\begin{table}[ht]
\footnotesize
\centering
\begin{tabular}{|c|c|c|c|}
\hline
             & Holders & Delegates & Proposals \\ \hline

     Uniswap & 368,193 &    27,805 &        39 \\ \hline
    Compound & 208,049 &     4,807 &       147 \\ \hline
         ENS &  64,290 &    11,834 &        12 \\ \hline
     Gas DAO &  44,987 &       586 &         2 \\ \hline
     Gitcoin &  31,595 &     6,341 &        45 \\ \hline
  Ampleforth &  26,236 &       255 &        13 \\ \hline
         Fei &  14,117 &       347 &        86 \\ \hline
         Hop &  14,000 &     4,148 &         4 \\ \hline
      Strike &   9,931 &         4 &        29 \\ \hline
PoolTogether &   8,393 &       466 &        60 \\ \hline
Rari Capital &   7,163 &        17 &         9 \\ \hline
     Radicle &   6,527 &        75 &        11 \\ \hline
     Indexed &   5,388 &       345 &        23 \\ \hline
  Braintrust &   3,962 &        12 &         3 \\ \hline
        Idle &   3,780 &        61 &        31 \\ \hline
   Instadapp &   3,469 &        32 &         4 \\ \hline
        Silo &   3,197 &        84 &        37 \\ \hline
     Inverse\footnote{For Inverse, data is not included up to 31 Jan 2023, see Appendix \ref{appendix:daos}.}
             &   2,409 &       191 &        29 \\ \hline
       Euler &   2,327 &       928 &         0 \\ \hline
     Cryptex &   1,581 &        12 &         9 \\ \hline
     Babylon &   1,090 &        47 &        28 \\ \hline

\end{tabular}
\caption{State of the analyzed governance systems on 31 Jan 2023.}
\label{table:data_overview}
\end{table}

\section{Distribution of Voting Power}

We begin by studying the distribution of voting power in the DAOs.
We do so using two measures: the Gini coefficient and the Nakamoto coefficient.

The Gini coefficient, one of the most frequently used inequality measures, was first introduced in 1912 by Corrado Gini \cite{ceriani2012gini_origins}. Originally, the coefficient was used to examine income and wealth inequality within a geographical community (e.g. a nation).
Nonetheless, it can be used to measure the inequality in the distribution of any fungible good, in our case voting power.
Its values range from 0.0 which indicates perfect equality to 1.0 meaning the highest level of inequality (a single individual possesses everything).
In most countries, the Gini coefficient of the distribution of wealth lies between 0.7 and 0.85 \cite{cs2022wealth}.

The Nakamoto coefficient, on the other hand, measures how decentralized a system is by counting how many parties are needed to collectively take control of the system. It was first formally described by Balaji Srinivasan in 2017 \cite{srinivasan2017nakamoto}.
Applied to a DAO governance system, the Nakamoto coefficient is defined as the number of addresses which together hold more than 50\% of the voting power.

\begin{table}[ht]
\footnotesize
\centering
\begin{tabular}{|c|c|c|c|c|}
\hline
      &  \thead{Nakamoto \\ Holders} &  \thead{Gini \\ Holders} &  \thead{Nakamoto \\ Delegates} &  \thead{Gini \\ Delegates} \\ \hline

         ENS &                94 &         0.914 &                  19 &           0.938 \\ \hline
     Gitcoin &                42 &         0.991 &                  10 &           0.993 \\ \hline
     Uniswap &                30 &         0.992 &                  11 &           0.999 \\ \hline
         Hop &                30 &         0.902 &                   6 &           0.967 \\ \hline
    Compound &                25 &         0.996 &                   6 &           0.996 \\ \hline
PoolTogether &                21 &         0.965 &                   7 &           0.949 \\ \hline
     Indexed &                16 &         0.935 &                   5 &           0.923 \\ \hline
Rari Capital &                15 &         0.918 &                   2 &           0.772 \\ \hline
     Babylon &                13 &         0.937 &                   5 &           0.731 \\ \hline
     Gas DAO &                12 &         0.888 &                   3 &           0.935 \\ \hline
  Braintrust &                11 &         0.966 &                   1 &           0.875 \\ \hline
        Silo &                11 &         0.964 &                   3 &           0.909 \\ \hline
  Ampleforth &                 9 &         0.984 &                   3 &           0.966 \\ \hline
        Idle &                 7 &         0.960 &                   2 &           0.900 \\ \hline
         Fei &                 7 &         0.975 &                  12 &           0.906 \\ \hline
     Radicle &                 6 &         0.990 &                   2 &           0.934 \\ \hline
     Cryptex &                 6 &         0.981 &                   2 &           0.637 \\ \hline
   Instadapp &                 2 &         0.979 &                   2 &           0.784 \\ \hline
     Inverse &                 2 &         0.944 &                   2 &           0.937 \\ \hline
       Euler &                 2 &         0.984 &                   3 &           0.980 \\ \hline
      Strike &                 1 &         1.000 &                   1 &           0.667 \\ \hline

\end{tabular}
\caption{Nakamoto and Gini coefficients of the distribution of voting power among token holders and among delegates at block 16,530,000 (31 Jan 2023).}
\label{tbl:gini_nakamoto}
\end{table}

When analyzing the distribution of power in DAOs with delegative token governance, there are two relevant distributions to consider: the distribution of governance tokens among token holders, and the distribution of voting power among delegates (i.e.\ the amount of tokens delegated to them by holders).
For both these distributions, Table \ref{tbl:gini_nakamoto} shows the Gini and Nakamoto coefficients for all analyzed DAOs on 31 Jan 2023.

With very few exceptions, the Gini coefficients are close to 1.0 indicating a highly unequal distribution of voting power. 
Furthermore, we observe this inequality remaining high over the whole observation period.

Regarding the values of the Nakamoto coefficient, it is notable how low they are across the board: Except for four projects, all DAOs have single digit Nakamoto coefficients for the distribution of voting power among delegates. This means that less than 10 addresses can take full control of the governance system and pass any decision they want. For half of the analyzed DAOs, the Nakamoto coefficient is even no larger than 3!

\begin{figure}[ht]
    \centering
    \includegraphics[width=0.80\textwidth]{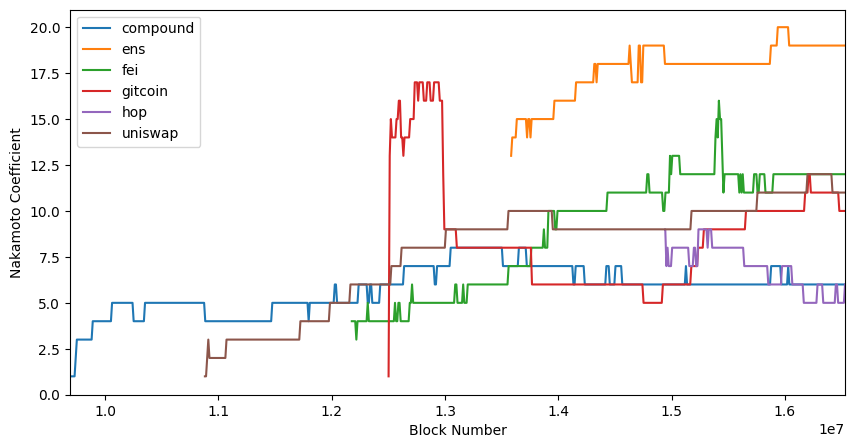}
    \caption{Nakamoto coefficients of voting power held by delegates for six selected DAOs between block 9,690,000 (17 Mar 2020) and 16,530,000 (31 Jan 2023).}
    \label{fig:nakamoto}
\end{figure}

Overall, our analysis shows that there is a very high degree of centralization of voting power in current DAO governance systems.
There is however a slight trend towards decentralization over time, as the increasing Nakamoto coefficients for delegates in Figure \ref{fig:nakamoto} show.
Nonetheless, these findings put a question mark behind the D in DAO.

\section{Structure of Voting Power Delegation}\label{sec:communityDelegates}

Besides the pure amount of voting power held by delegates, as analyzed in the previous section, another relevant aspect to the nature of a governance system is who the delegates are representing.
Do they tend to represent large token holders (possibly themselves) or a group of community members?

In the following, we examine the structure of the delegations of voting power.
To that end, we use the distinction between \emph{single holder delegates} and \emph{community delegates} introduced in \cite{fritsch2022daos}. Single holder delegates are delegates who receive more than 50\% of the tokens delegated to them from a single token holder. A delegate receiving less than 50\% of delegated tokens from a single holder is called a community delegate.
The idea behind this definition is that single holder delegates mainly represent a single holder (and their interests). This includes the case of large token holder delegating to themselves due to the need to delegate before voting.

In the following, we measure the share of votes held by all community delegates.
Governance systems with a large share of community delegates can be said to somewhat resemble a representative (parliamentary) system with a community electing representatives.
If this share is low however, this indicates that the delegation part of the governance system is not being utilized to a large extent.
Such governance systems then mainly feature direct representation of the interests of large token holders.

\begin{figure}[ht]
    \centering
    \includegraphics[width=0.80\textwidth]{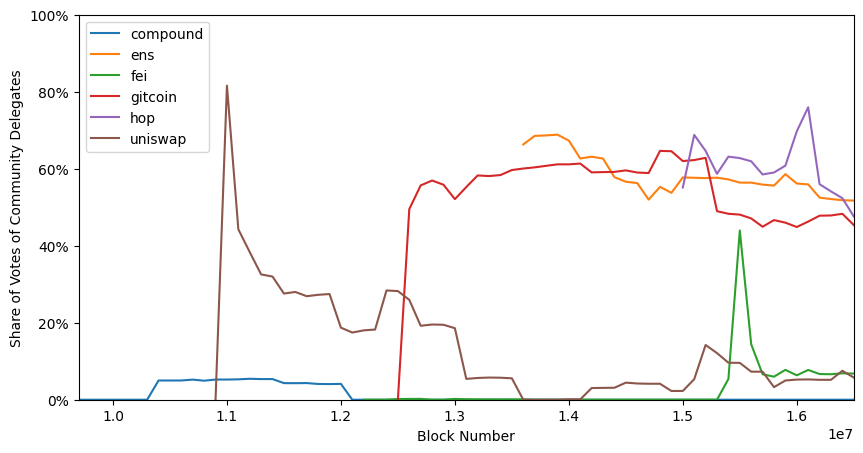}
    \caption{Share of voting power held by community delegates (i.e.\ delegates who are not delegated more than 50\% of their votes by a single holder) for six selected DAOs between block 9,690,000 (17 Mar 2020) and 16,530,000 (31 Jan 2023).}
    \label{fig:community_delegates}
\end{figure}

In Figure \ref{fig:community_delegates}, we see that for ENS, Gitcoin and Hop about half of all votes are in the hands of community delegates.
For Compound, Fei and Uniswap on the other hand, the vote share of community delegates is low at about 10\% or less. This is also the case for most other DAOs in the dataset as Table \ref{table:governance_overview} shows. For many of them, almost all voting power is held by delegates who mainly represent a single holder.

Again, our results show a very low degree of decentralization.
Furthermore, they call into question the necessity of a delegation system, since this comes with significant cost as we will show in following sections.

\section{Governance Participation}

The governance participation rate can be defined in different ways, namely among tokens, delegates and voting power.
The participation rate of the token holders is defined as the proportion of token holders voting out of the total number of token holders at the time of the vote. Accordingly, the participation rate among the delegates is the proportion of delegates voting among the total number of delegates.
Furthermore, we consider the indirect participation rate of holder, i.e. the proportion of holders involved in a vote (voting directly or represented by a delegate) among all holders.
Finally, the participation rate of voting power is the number of governance tokens voting relative to the total number of governance tokens delegated at the time of voting.

\begin{figure}[ht]
\centering
\begin{subfigure}[t]{.5\textwidth}
  \centering
    \includegraphics[width=1.0\columnwidth]{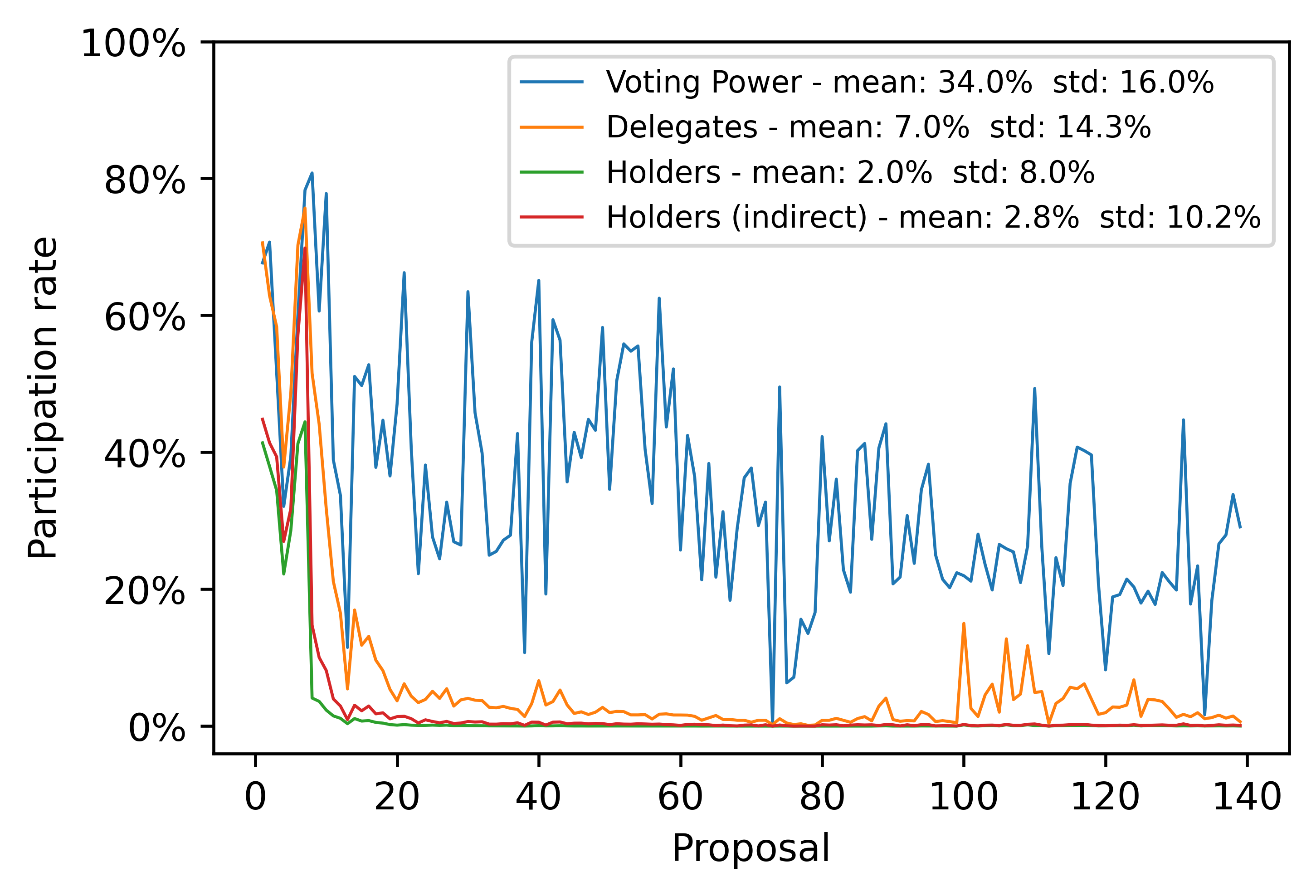}
    \caption{Compound}
    \label{fig:participationrate_compound}
\end{subfigure}%
\begin{subfigure}[t]{.5\textwidth}
  \centering
  \includegraphics[width=1.0\columnwidth]{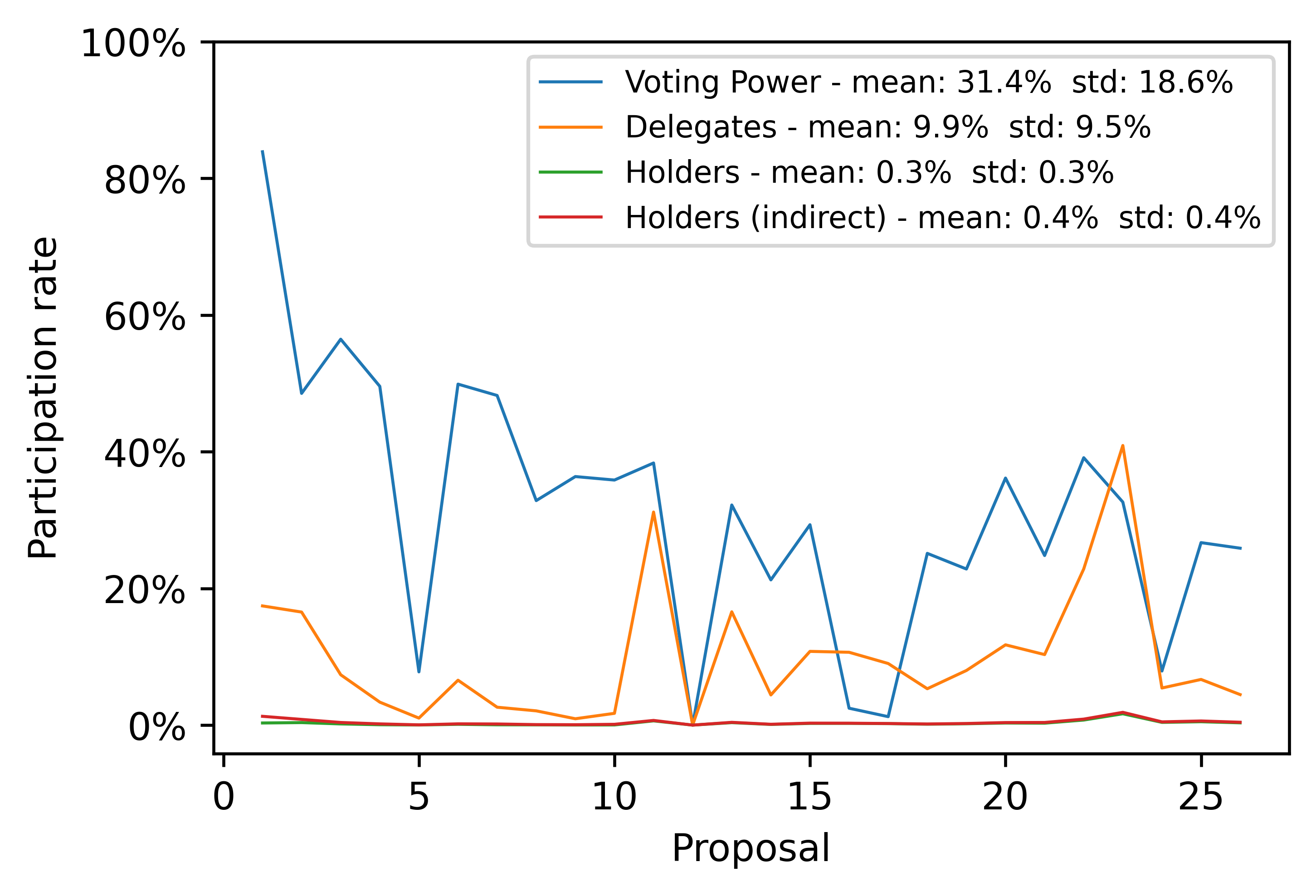}
    \caption{Uniswap}
    \label{fig:participationrate_uniswap}
\end{subfigure}
\begin{subfigure}[t]{.5\textwidth}
    \centering
    \includegraphics[width=1.0\columnwidth]{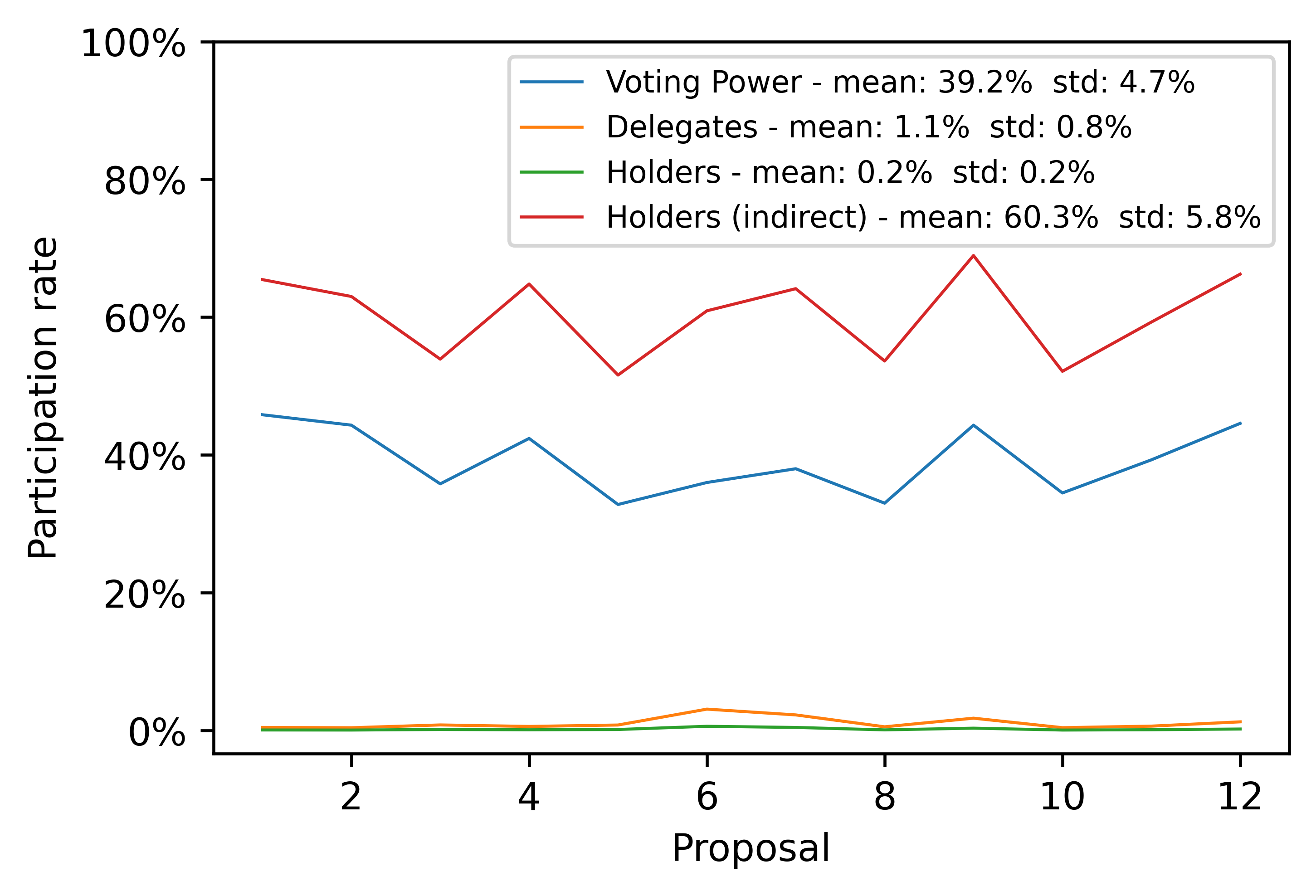}
    \caption{ENS}
    \label{fig:participationrate_ens}
\end{subfigure}%
\begin{subfigure}[t]{.5\textwidth}
    \centering
    \includegraphics[width=1.0\columnwidth]{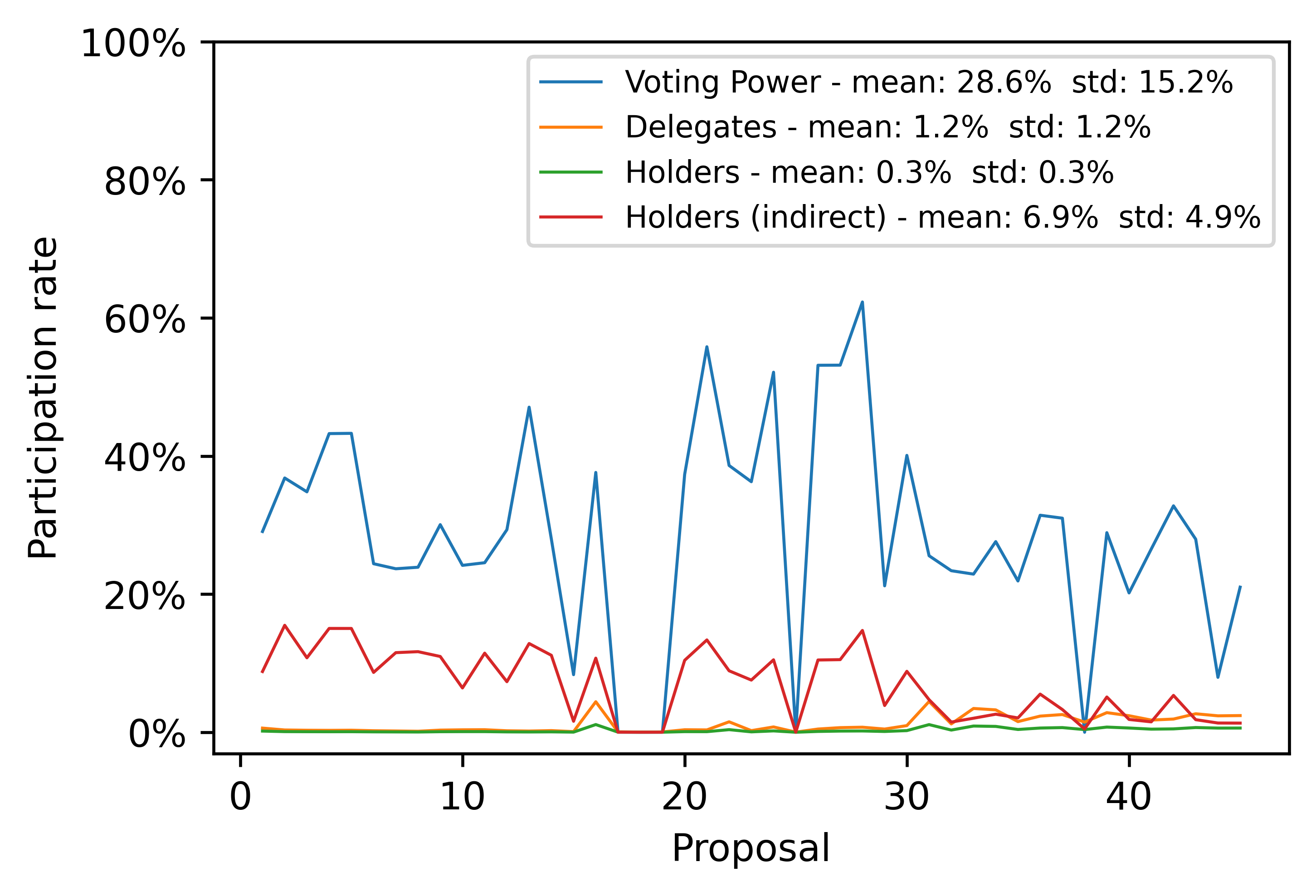}
    \caption{Gitcoin}
    \label{fig:participationrate_gitcoin}
\end{subfigure}
\caption{Participation rates among holders, delegates, and voting power.}
\label{fig:participation}
\end{figure}

Figure \ref{fig:participation} shows the participation rates for Compound, Uniswap, ENS and Gitcoin.
The participation rate of the voting power is typically the higher than the rate for holder and delegates. This means that those voters with particularly high voting power are more active. This is not surprising since their costs relative to exercised voting power are lower. Moreover, delegates with high voting power either own many tokens themselves -- and are thus strongly affected by the proposals voted on -- or have many tokens delegated to them -- and with that have a certain (moral) responsibility to vote.
The participation rate of token holders is particularly low for all DAOs. The initially high value for Compound can be explained by the low number of token holders at the beginning of governance. The fluctuations in the participation rates also show that the interest in the proposals depends on their content. 

The governances of ENS (Figure \ref{fig:participationrate_ens}) and Gitcoin (Figure \ref{fig:participationrate_gitcoin}) show a comparatively low delegate participation rate. This phenomenon can be explained by the fact that these two protocols required delegating when claiming their airdrop.
For Compound and Uniswap, only those interested in participating in proposals have to delegate. It turns out that for ENS and Gitcoin, many delegates who were created during the airdrop do not participate in governance.

On the other hand, Figure \ref{fig:participationrate_ens} shows a particularly large amount of ENS token holders (on average about 60\%) being represented during votes (either directly or indirectly by their delegate voting for them).
This is a positive indication of a working governance system, and a pro argument for requiring delegations when for requiring delegations when claiming tokens.

\section{Pointless Governance Transactions}

The raw numbers of votes and delegations, as analyzed in the previous section, often do not show the whole picture of how active a DAO really is.
Many transactions may simply be made by addresses hunting future airdrops, and some may be straight up mistakes.

We define \emph{pointless transactions} as transactions that have no discernible use, and are most likely the result of an error by a user. In particular, we have found three types of pointless transactions.
\begin{itemize}
 \item \textbf{Pointless transfers} are transactions that transfer zero tokens, or where recipient = sender.
 \item \textbf{Pointless votes} are votes cast by accounts holding no voting power. 
 \item \textbf{Pointless delegations}, are delegations for which the new delegate is equal to the old delegate.
\end{itemize} 

We consider the number of pointless transactions as a proxy to measure a community's \emph{maturity}. We analyze the share of pointless transactions and assume that a decrease would hint at the community getting more accustomed to the functionality of the respective governance protocols.  

\begin{figure}
\centering
\begin{subfigure}[t]{.5\textwidth}
 \centering
     \includegraphics[width=1.0\columnwidth]{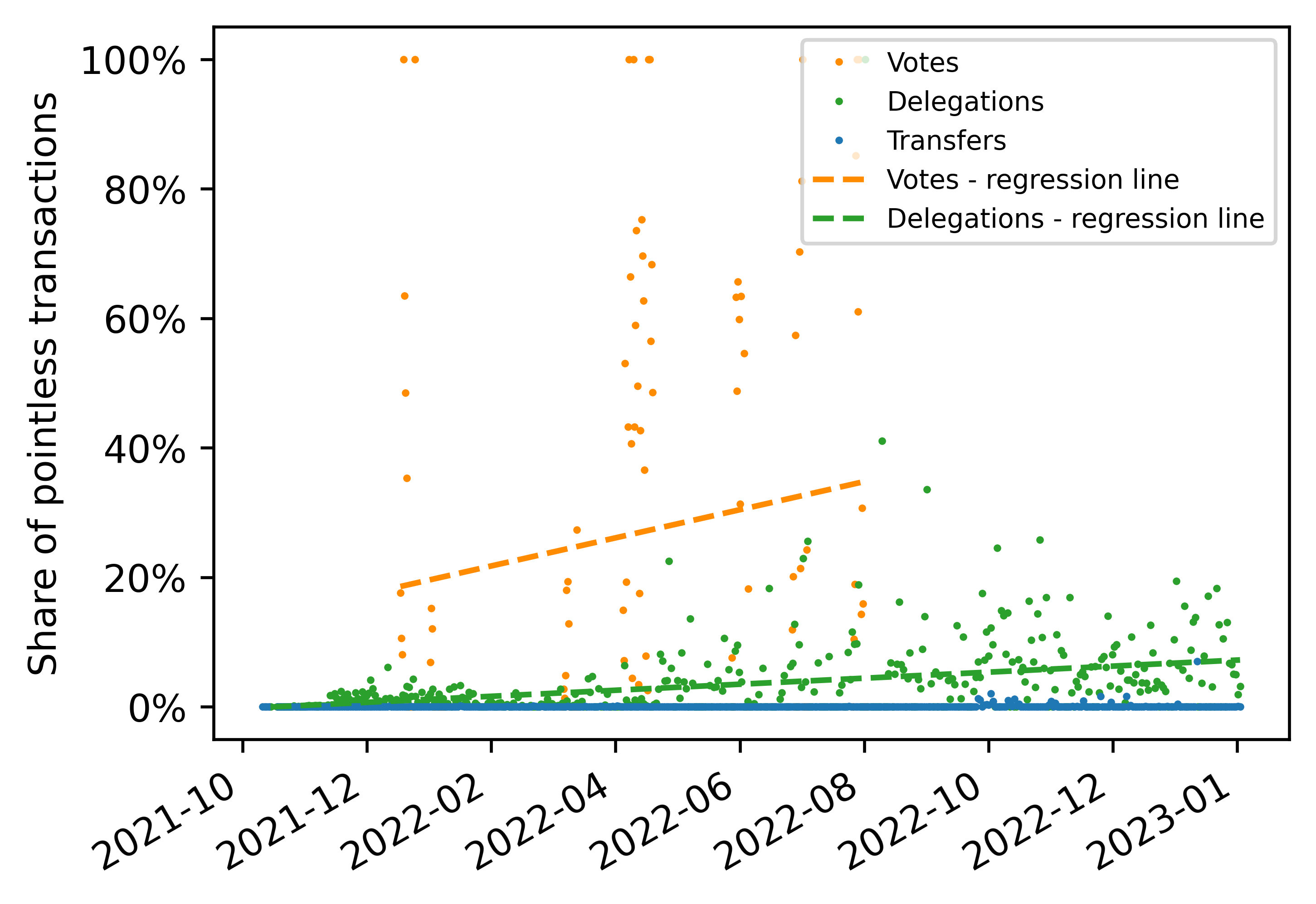}
    \caption{ENS}
    \label{fig:share_useless_transactions_cost_ens}
   \end{subfigure}%
\begin{subfigure}[t]{.5\textwidth}
  \centering
  \includegraphics[width=1.0\columnwidth]{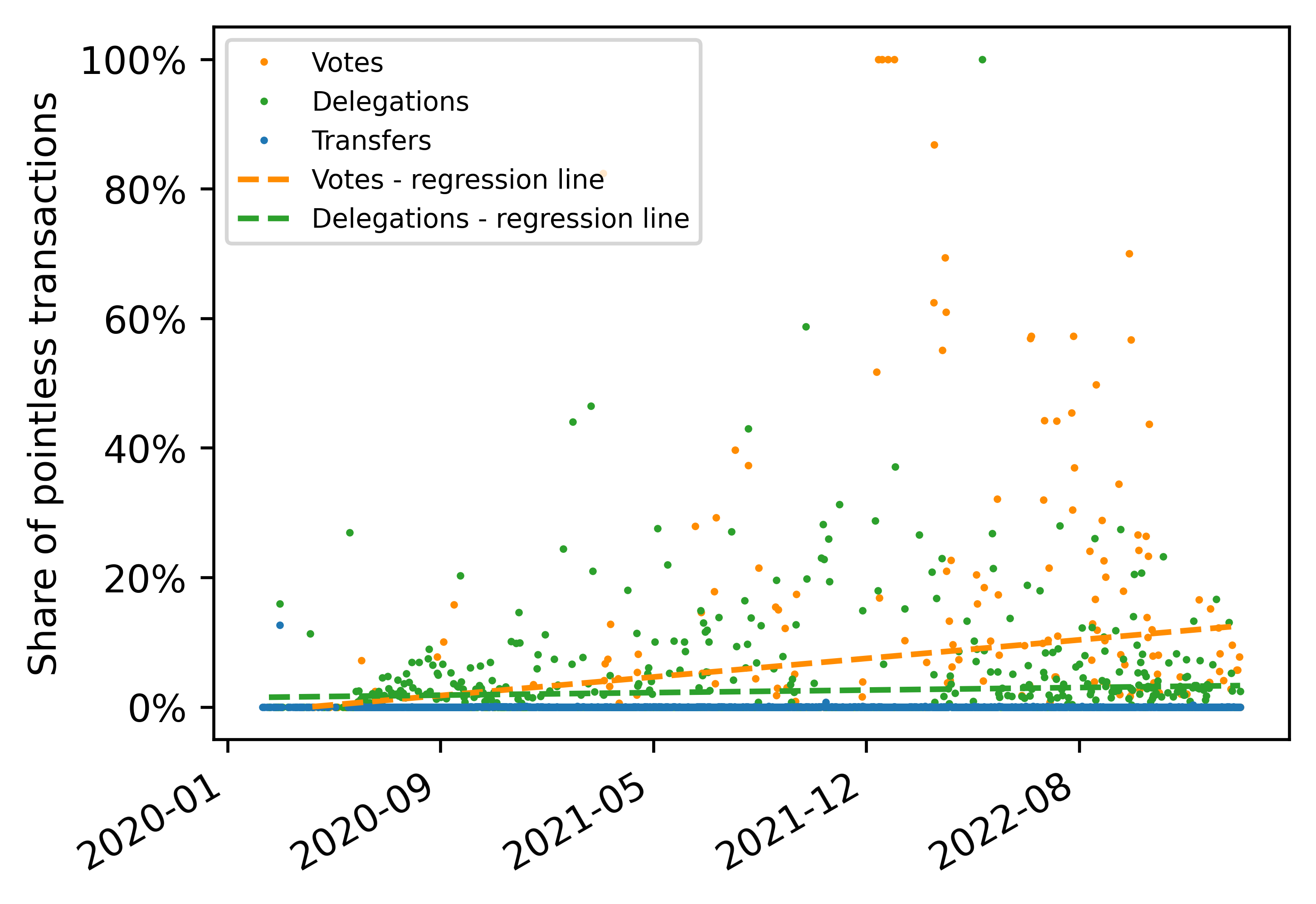}
    \caption{Compound}
    \label{fig:share_useless_transactions_cost_compound}
\end{subfigure}
\caption{Proportion of pointless transactions in all transactions per day.}
 \label{fig:share_useless_transactions}
\end{figure}

Figure \ref{fig:share_useless_transactions} shows the share of pointless transactions per day. Overall we observe that pointless transfers are rare. We assume that most users already have experience with the use of the transfers function or perform the transfer with an interface that indicates useless transfers. Astonishingly, the proportion of pointless votes, and to a lesser degree pointless delegations, is exceptionally high. For ENS for instance, on average almost 30\% of all votes per day are pointless. We suspect that many users do not realize that they must delegate their voting power to themselves before being allowed to vote. It is also remarkable that the proportion of pointless delegations and votes per day increases over time for ENS and Compound. Therefore, we conclude that DAOs are still very much in their infancy and have not reached maturity.

Table \ref{table:governance_overview} shows the proportion of useless transactions when considering votes and delegations for each DAO. In general, this proportion is shockingly high with most DAOs having more than 10\% useless votes and delegations, many even more than 20\%. Further note that we used a very conservative definition of pointless transactions. Indeed, for Uniswap for instance, 88\% of votes cast have a voting power below 10 tokens, and a staggering 47\% of votes have a voting power below 1 token, while 2.5 and 40 million tokens are required to submit and pass a proposal respectively.

Note that overall ENS actually only has a few pointless transactions (<2\%). This does not contradict Figure \ref{fig:share_useless_transactions} since the majority of governance transactions occurred shortly after its airdrop, when the percentage of pointless transactions was low.

\section{Monetary Price of Governance}\label{sec:costOfGovernance}

In this section, we analyze the cost of performing governance on-chain. For each transaction carried out on blockchains such as Ethereum, a fee must be paid. The amount of the fee depends on the computational effort of the respective transactions (measured in units of gas), as well as the current price for unit of gas (which depends on the demand for block space).

To compute the monetary price of governance, we first consider transactions that handle voting, delegation voting power, and creating proposals. We define the \emph{price of governance transactions} to be the sum of fees paid for these three types of transactions. Note that in some instances, multiple actions are combined into a single transaction. In such cases, we take special care only to include costs related to the governance actions in the transaction. A more thorough description on how this is achieved is detailed in Appendix \ref{appendix:costs}.

Both the gas price in ETH and the price of ETH in USD are subject to strong fluctuations. Therefore, we consider the ETH price as observed on Etherscan on the day of a transaction, and report the cost of governance in USD.

\subsection{Price of Governance Transactions}

\begin{figure}[ht]
\centering
\begin{subfigure}[t]{.5\textwidth}
    \centering
    \includegraphics[width=1.0\columnwidth]{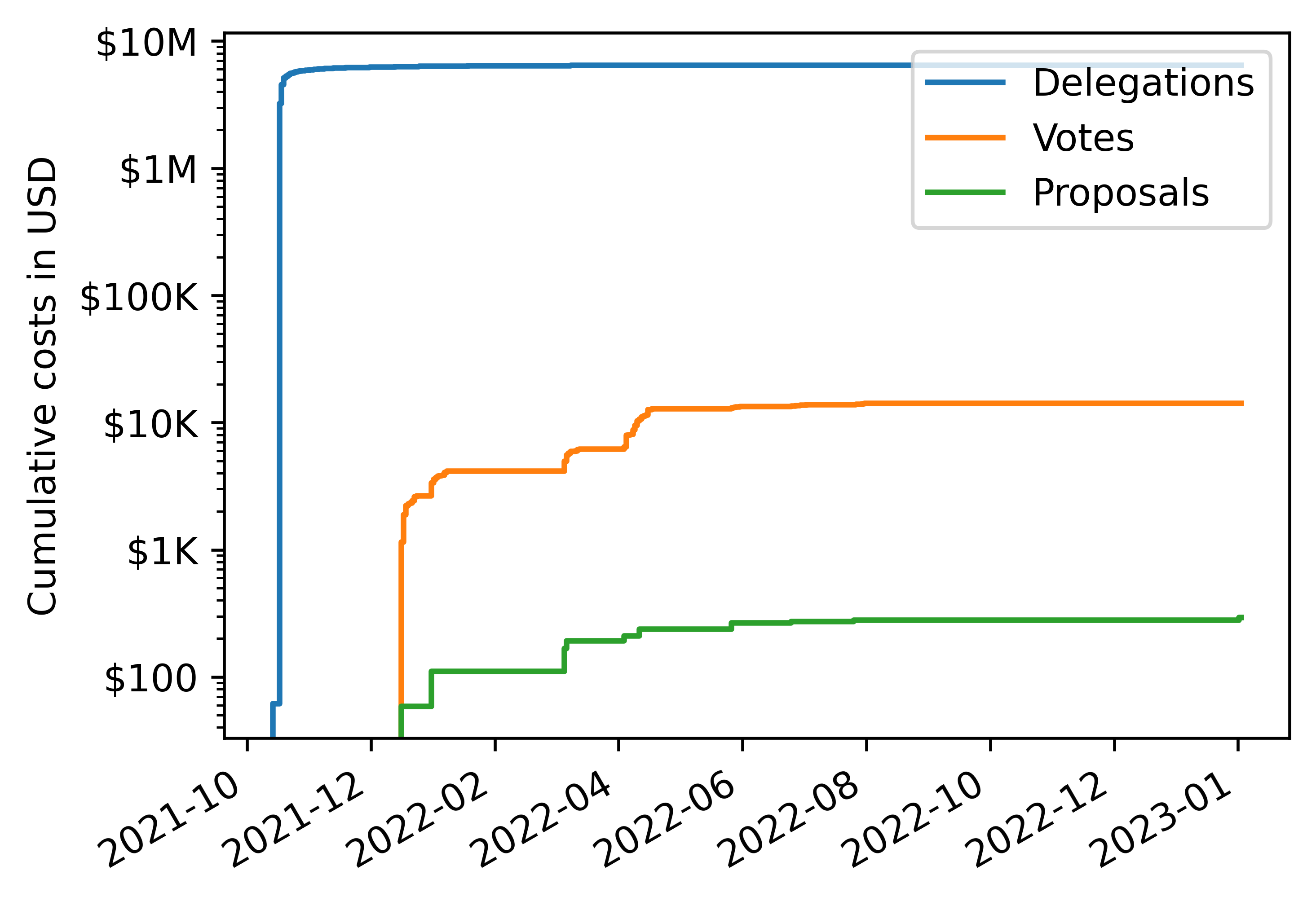}
    \caption{ENS (note the logarithmic y-axis)}
    \label{fig:transactions_cost_ens}
\end{subfigure}%
\begin{subfigure}[t]{.5\textwidth}
    \centering
    \includegraphics[width=1.0\columnwidth]{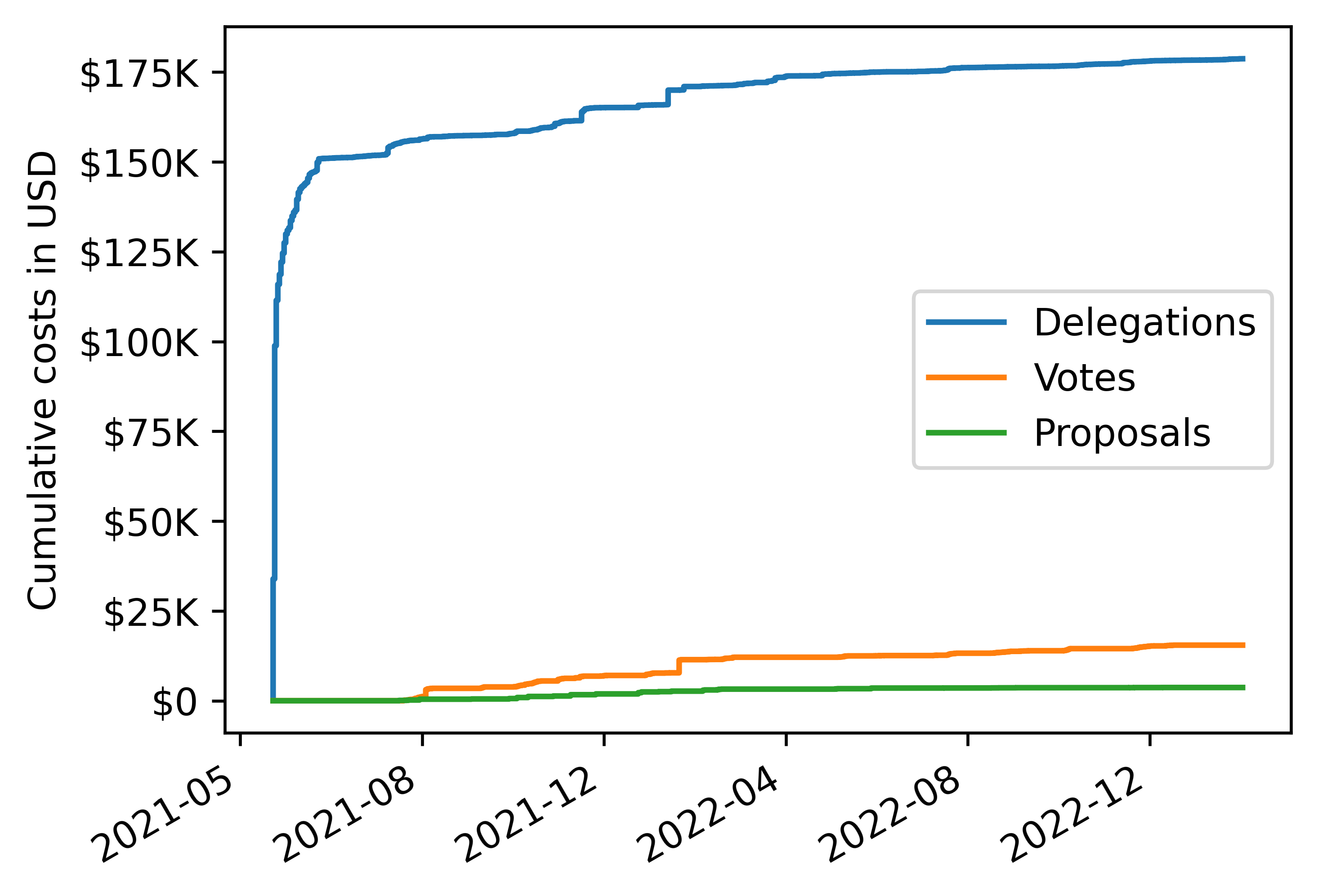}
    \caption{Gitcoin}
    \label{fig:transactions_cost_gitcoin}
\end{subfigure}
\begin{subfigure}[t]{.5\textwidth}
    \centering
    \includegraphics[width=1.0\columnwidth]{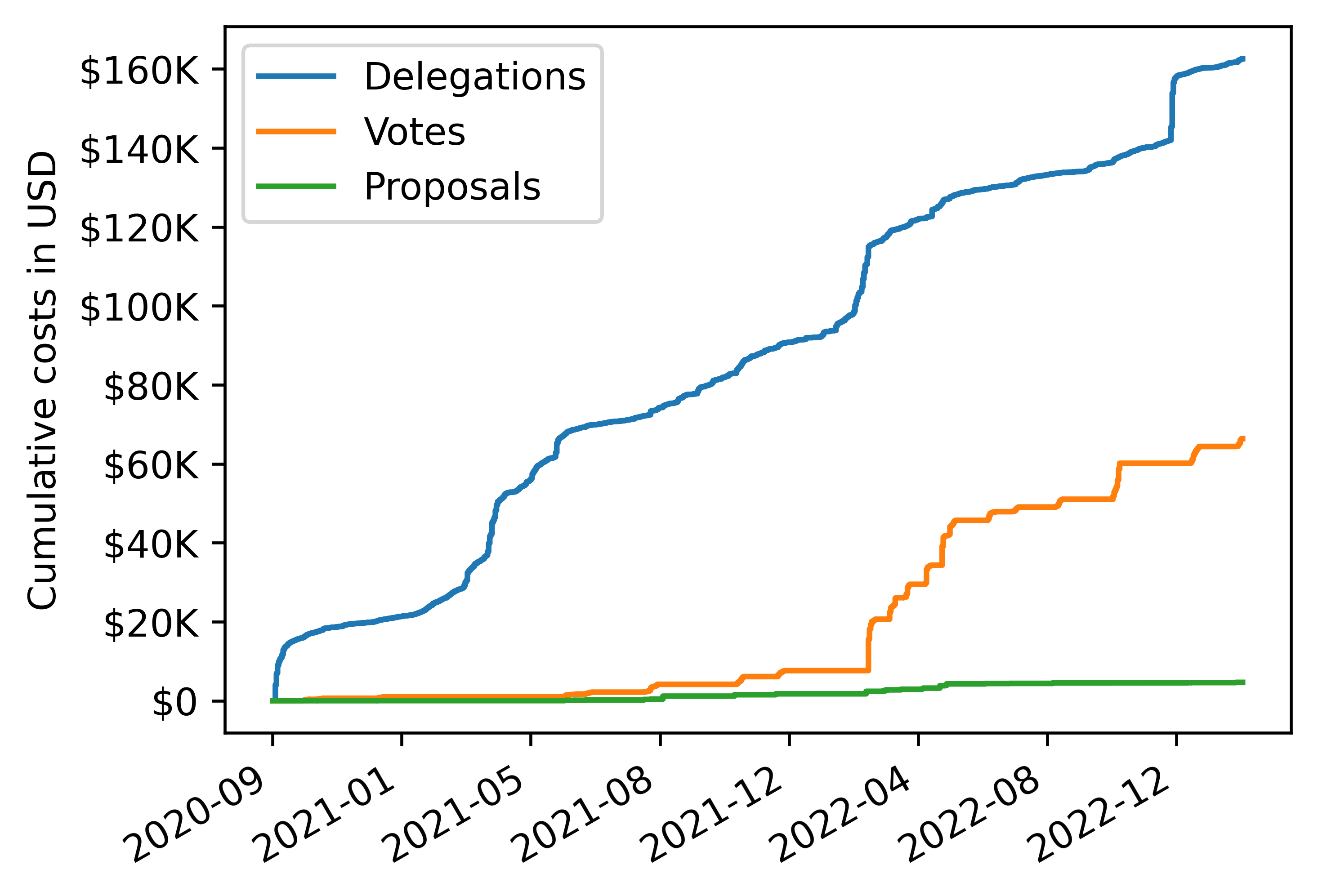}
    \caption{Uniswap}
    \label{fig:transactions_cost_uniswap}
\end{subfigure}%
\begin{subfigure}[t]{.5\textwidth}
    \centering
    \includegraphics[width=1.0\columnwidth]{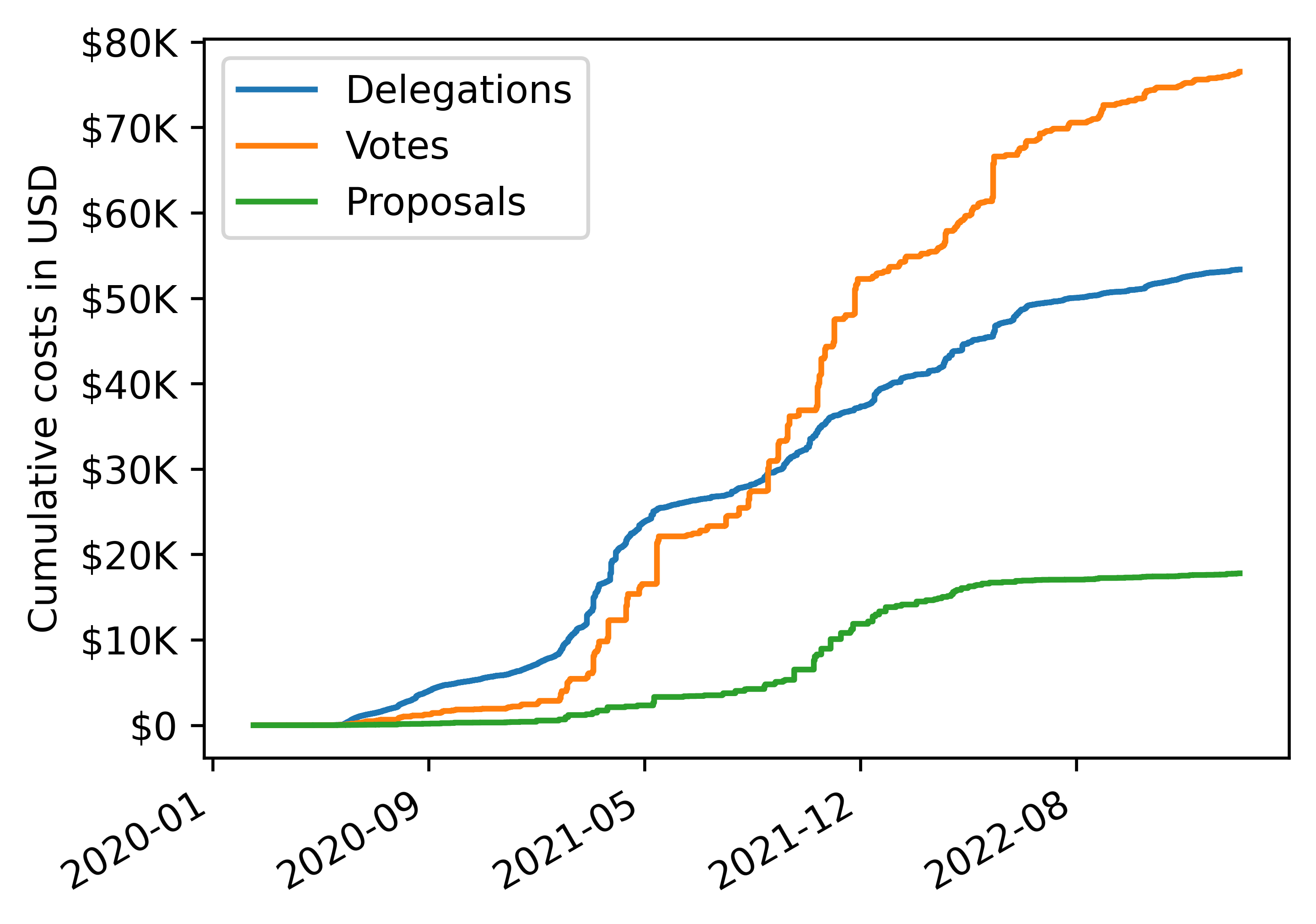}
    \caption{Compound}
    \label{fig:transactions_cost_compound}
\end{subfigure}
\caption{Cost of governance: Transaction costs of delegations, votes and proposals.}
\label{fig:cost_of_governance}
\end{figure}

Transactions that create proposals are much larger than delegation or voting transactions, and are thus much more costly.  However, the number of proposals created is typically much smaller than the number of delegations and votes cast. Thus, as can be seen in Figure \ref{fig:cost_of_governance}, costs are usually dominated by delegations. One exception is Compound, the longest running DAO in the dataset, where voting costs have overtaken delegation costs.

\begin{figure}
\centering
\begin{subfigure}[t]{.5\textwidth}
 \centering
     \includegraphics[width=1.0\columnwidth]{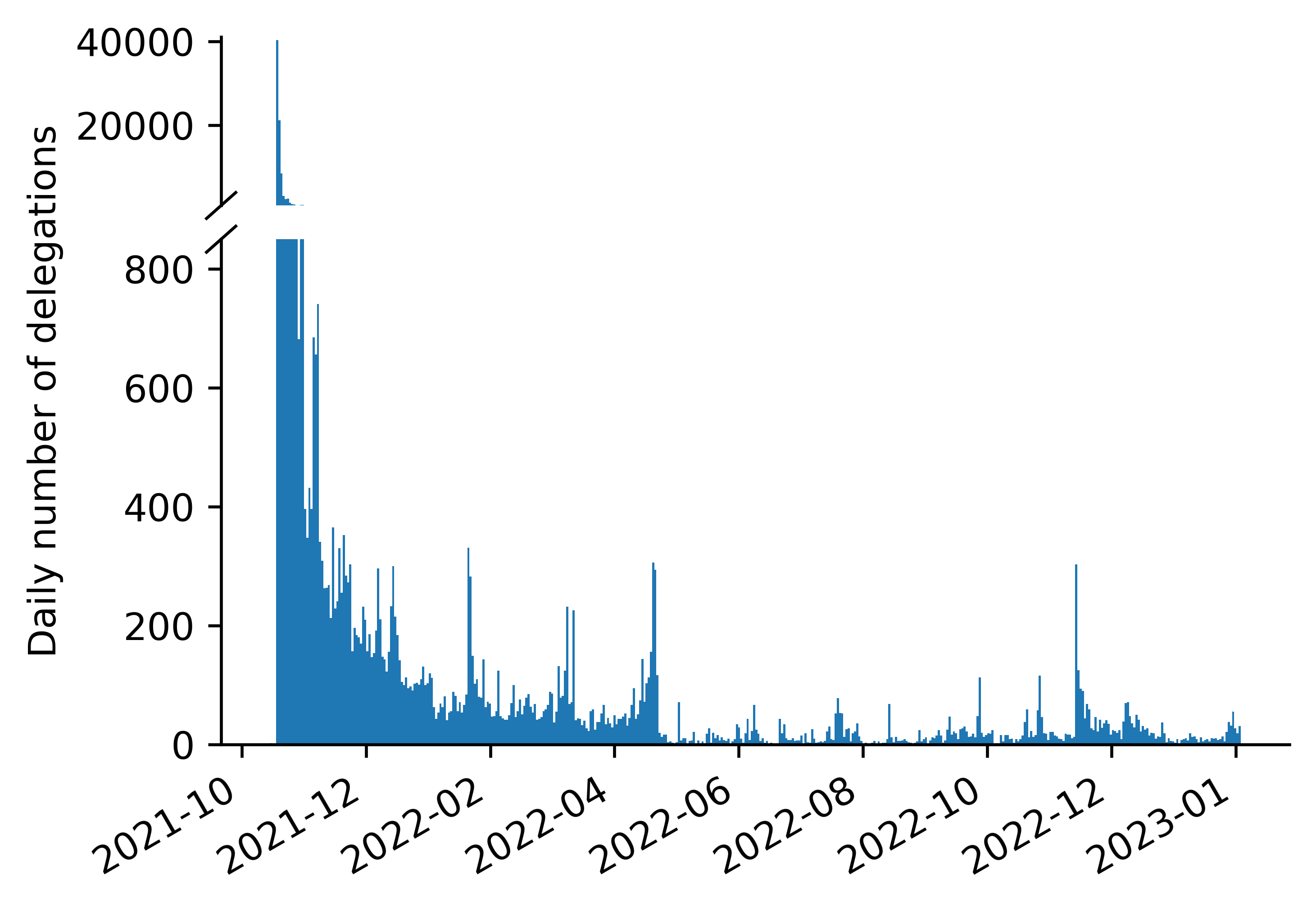}
    \caption{ENS}
    \label{fig:delegations_ens}
   \end{subfigure}%
\begin{subfigure}[t]{.5\textwidth}
  \centering
  \includegraphics[width=1.0\columnwidth]{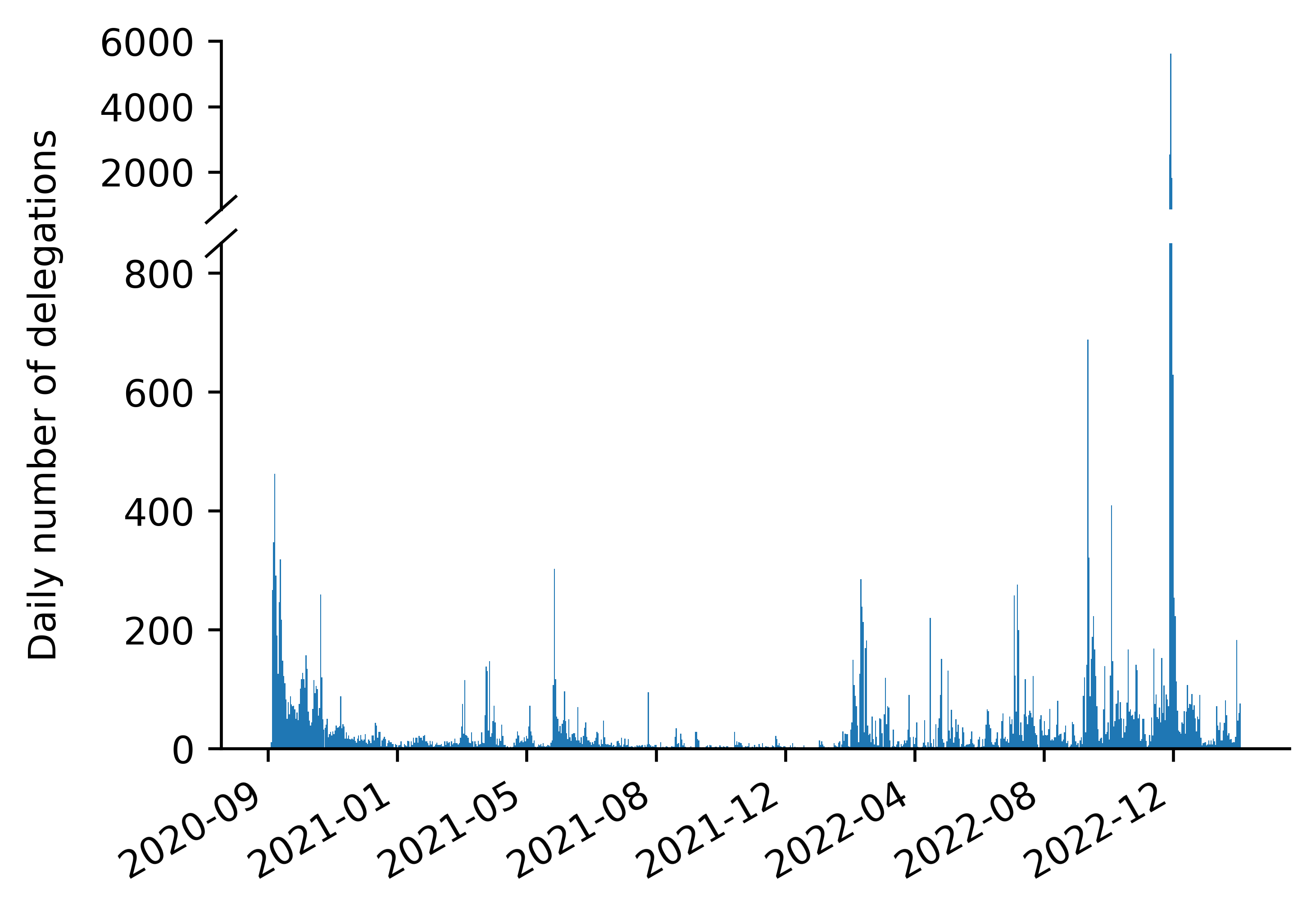}
    \caption{Uniswap}
    \label{fig:delegations_uniswap}
\end{subfigure}
\caption{Number of delegations per day.}
\end{figure}

The total governance transaction costs for all DAOs are listed in Table \ref{table:governance_overview}.
In particular, the table shows that ENS has exceptionally high costs of the governance with about \$3.5 million (costs are below \$300.000 for all other DAOs). As Figure \ref{fig:transactions_cost_ens}) shows, this is due to particularly high delegation costs following the launch of the governance system.
When the previously mentioned ENS Airdrop took place in November 2021, the Ether price in USD was exceptionally high, and so was the price of a unit of gas. As virtually all delegations took place during that time (see Figure \ref{fig:delegations_ens}) due to the requirement to delegate when claiming the airdrop, this resulted in extensive amounts of fees being paid for governance transactions.
On the other hand, Gitcoin, a protocol that also enforced delegations upon claiming an airdrop, started in May 2021, and thus benefitted from lower fees (see Figure \ref{fig:transactions_cost_gitcoin}). Of course absolutely speaking, Gitcoin also has much fewer token holders than ENS.

In contrast to ENS and Gitcoin, Uniswap and other protocols did not implement a such delegation requirement. This lead to delegations and costs being much more spread out over time. Nonetheless, Figure \ref{fig:delegations_uniswap} shows a large peak between the 29 Nov 2022 and the 1 Dec 2022 for Uniswap. We trace these delegations back to newly created accounts with very low token balances and similar transaction patterns. We hypothesize that most of the observed activity in this time period comes from one or more airdrop hunters trying to set up wallets for airdrop farming.

In terms of architecture, we conclude the mandatory delegation during the airdrop can be particularly cost-intensive. However, the high costs might be justified by the hope of achieving higher participation and a more even distribution of voting power. The primary function of governance tokens is in most cases participation in governance. In this respect, it can also be argued that a token holder who does not delegate does not fulfil the actual purpose of the token.

\begin{figure}
\centering
\begin{subfigure}[t]{.5\textwidth}
    \centering
    \includegraphics[width=1.0\columnwidth]{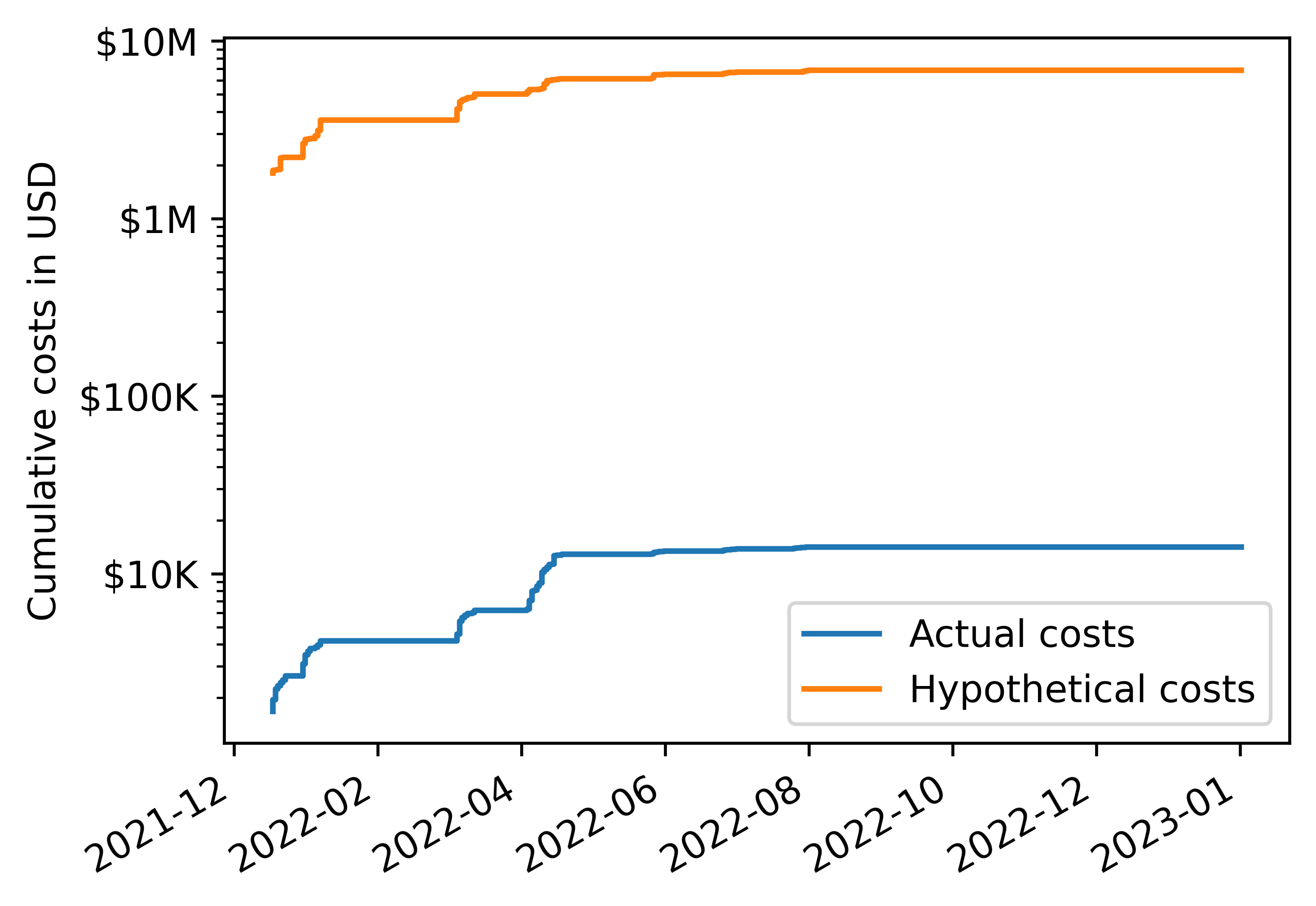}
    \caption{ENS}
    \label{fig:cost_votes_hypo_ens}
   \end{subfigure}%
\begin{subfigure}[t]{.5\textwidth}
    \centering
    \includegraphics[width=1.0\columnwidth]{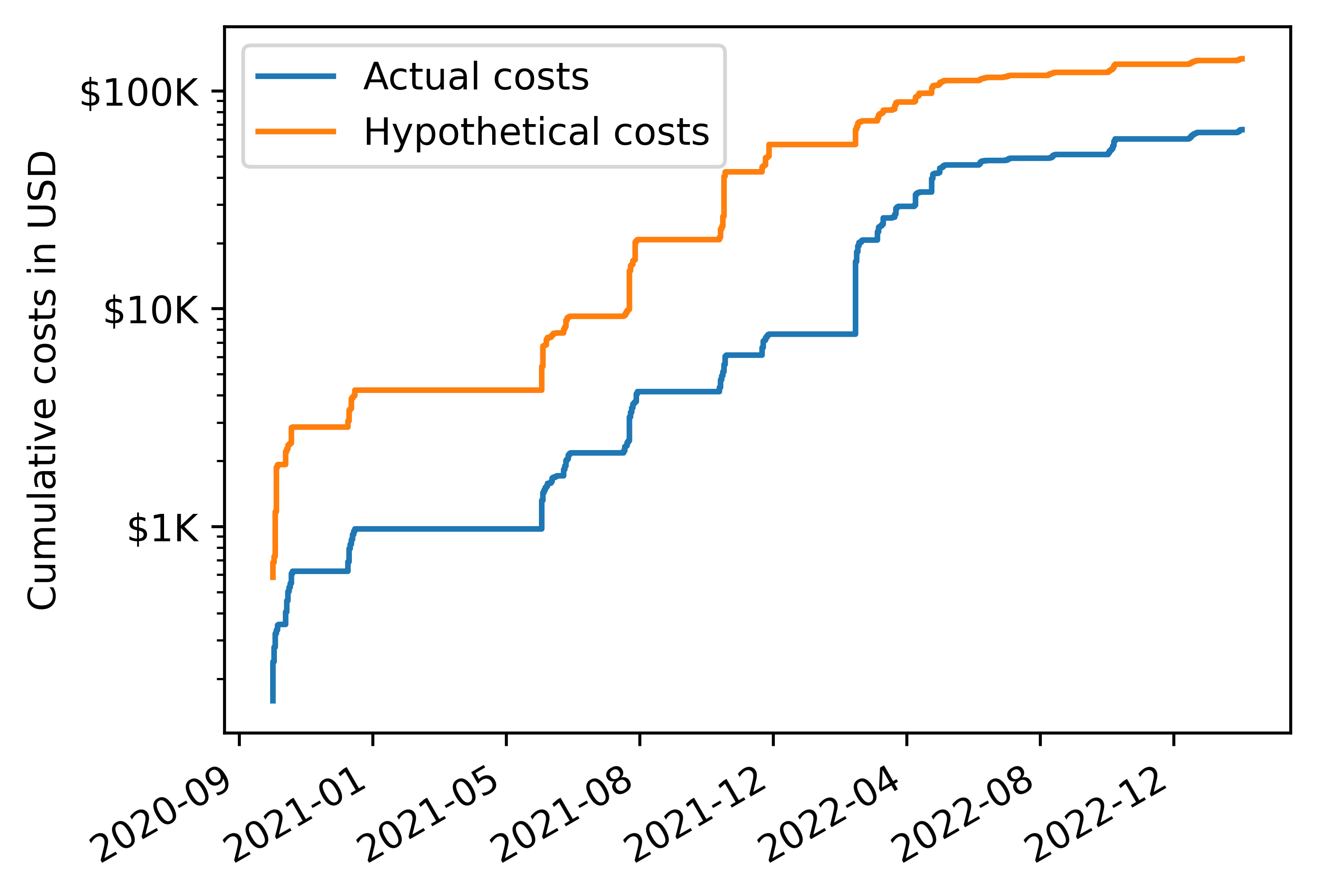}
    \caption{Uniswap}
    \label{fig:cost_votes_hypo_uniswap}
\end{subfigure}
\caption{Hypothetical costs of vote transactions if no delegation mechanism existed and all participation holders needed to vote individually.}
\label{fig:cost_votes_hypo}
\end{figure}

After quantifying the costs of delegations, we quantify the cost savings they result in.
Since the cost of a vote transaction is independent of the number of token holders a delegate represents, delegations lead to savings in transaction fees for voting. Figure \ref{fig:cost_votes_hypo} shows the cost for vote transactions that would arise if, all other things being equal, there were no delegations and all token holders represented by delegates voted individually.
Comparing Figures \ref{fig:transactions_cost_ens} and \ref{fig:cost_votes_hypo_ens}, we find that for ENS, the savings from delegations actually about make up for their costs.
Finally, note that the savings from delegations are a lot smaller for Uniswap (Figure \ref{fig:cost_votes_hypo_uniswap}), in line with the observation that delegations being used less there (see Section \ref{sec:communityDelegates}).





\subsection{Price of Transfer Overhead}

A DAO's governance token often serves a dual purpose. On the one hand, as the name suggests, each token grants one voting right in the protocol's governance. On the other hand, tokens are traded as a monetary asset, and are often regarded as a way to participate in the success of a project, akin to a stock. Holding tokens might also generate revenue through their incorporation in the broader DeFi ecosystem, e.g.\ offering yield through staking, lending, or direct dividends from the protocol. 

Generally, one would expect that the free exchange of tokens and the associated costs can be regarded separately from the price of governance.
However, when a token offers the dual purpose of monetary asset and voting right, a hidden cost comes into play, that is not present for pure ERC-20 tokens.

The crucial insight is that each token transfer might change the voting power of delegates and thus requires additional smart contract logic, whose operation on the Ethereum blockchain we show to incur non-negligible additional costs. We explain how we compute these overhead costs in Appendix \ref{appendix:overhead_costs}.

\begin{figure}[ht]
\centering
\begin{subfigure}[t]{.5\textwidth}
\centering
    \includegraphics[width=1.0\columnwidth]{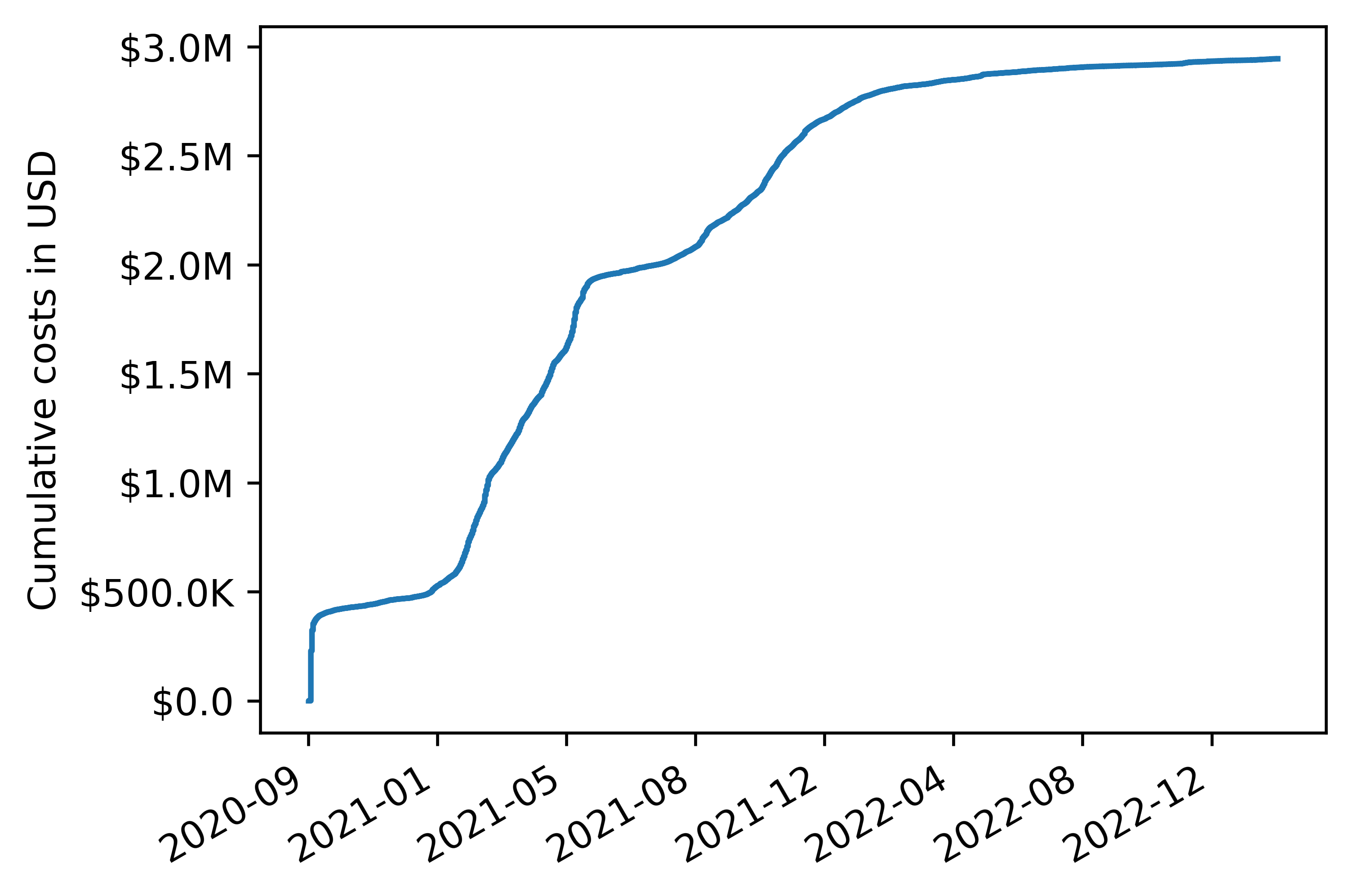}
    \caption{Uniswap}
    \label{fig:transfer_overhead_uniswap}
   \end{subfigure}%
\begin{subfigure}[t]{.5\textwidth}
  \centering
    \includegraphics[width=1.0\columnwidth]{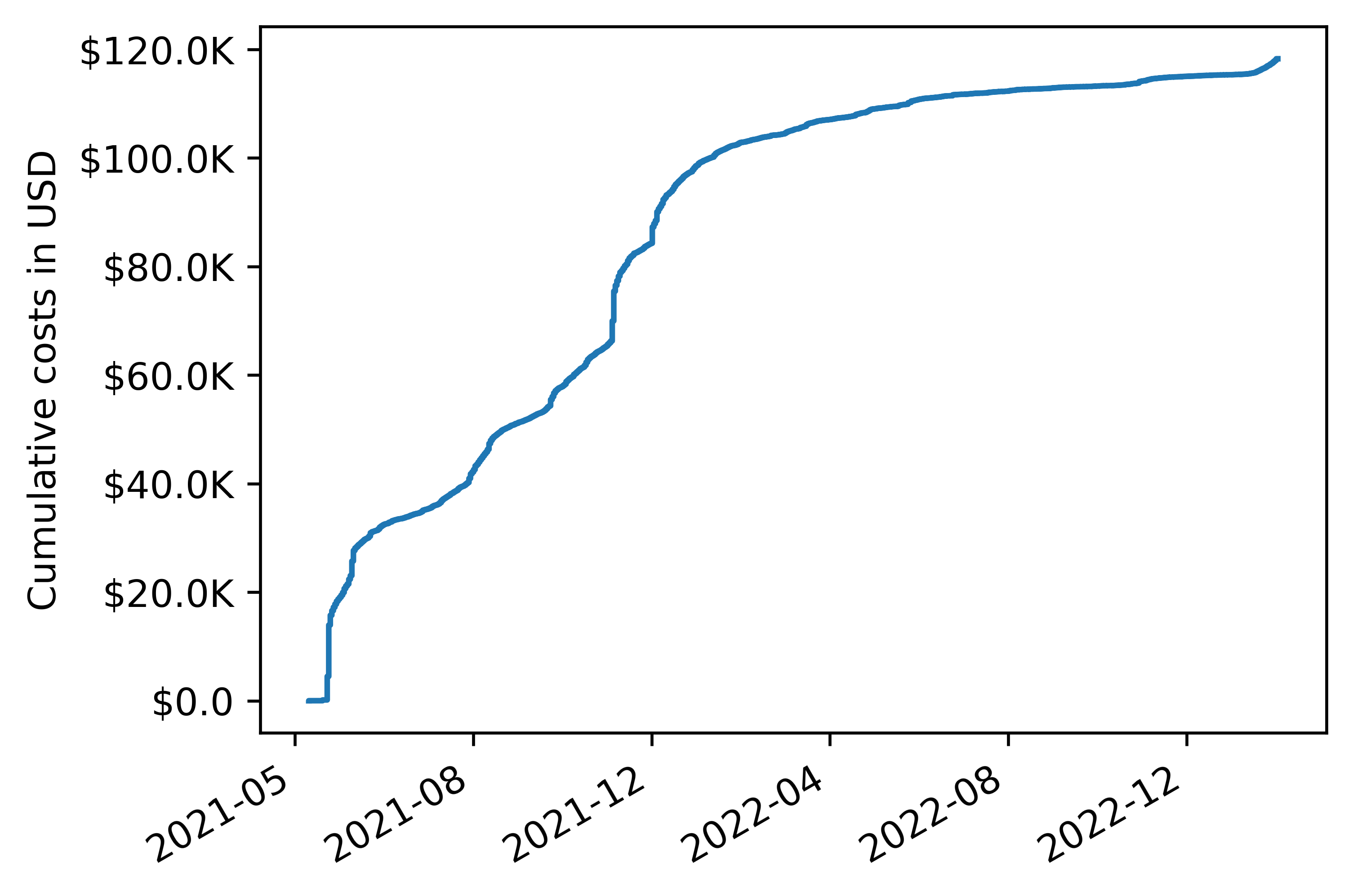}
    \caption{Gitcoin}
    \label{fig:transfer_overhead}
\end{subfigure}
\caption{Additional costs incurred by the modified transfer function.}
\end{figure}

Figure \ref{fig:transfer_overhead_uniswap} shows that for Uniswap for instance, the costs incurred by this seemingly small variation in the smart contract has cost users just shy of 3 Million USD, and thus dwarves the cost of direct governance transactions that amount to around 230'000 USD (see Figure \ref{fig:transactions_cost_uniswap}). 

The total cost of governance encompassing both the transfer overhead costs and cost of governance is shown in Table \ref{table:governance_overview}. Especially for projects generating low revenue, or projects that have very infrequent and inactive governance protocols, the current implementation might have to be questioned going forward.

\begin{table}[htb]
\footnotesize
\centering
\begin{tabular}{|c|c|c|c|c|c|}
\hline
             &  \thead{Vote Share of\\Community\\Delegates} &  \thead{Participation\\Rate of\\Voting Power} & \thead{Cost of\\Governance\\Transactions}  &  \thead{Total\\Cost of\\Governance} &  \thead{Proportion of\\Pointless\\Transactions} \\ \hline

  Ampleforth &           0.2\% &             77.4\% &                   \$9,303 &           \$248,139 &        15.9\% \\ \hline
     Babylon &           2.3\% &             33.8\% &                  \$13,864 &            \$22,772 &        44.9\% \\ \hline
  Braintrust &           0.0\% &             44.8\% &                   \$4,015 &            \$30,193 &        28.9\% \\ \hline
    Compound &           0.0\% &             32.2\% &                 \$147,659 &         \$1,221,492 &         9.8\% \\ \hline
     Cryptex &           0.0\% &             53.1\% &                   \$2,255 &            \$45,282 &         6.8\% \\ \hline
         ENS &          55.9\% &             39.2\% &               \$6,501,217 &         \$7,705,617 &         1.6\% \\ \hline
       Euler &           2.3\% &                N/A &                   \$6,832 &            \$11,333 &        15.7\% \\ \hline
         Fei &           0.0\% &             11.2\% &                  \$58,923 &           \$312,097 &        23.3\% \\ \hline
     Gas DAO &          38.6\% &             54.9\% &                  \$59,533 &           \$351,225 &        15.0\% \\ \hline
     Gitcoin &          43.7\% &             28.6\% &                 \$197,841 &           \$316,112 &         9.2\% \\ \hline
         Hop &          47.6\% &             43.8\% &                 \$217,902 &           \$238,351 &         2.7\% \\ \hline
        Idle &           0.0\% &             36.5\% &                  \$15,335 &            \$89,350 &        13.2\% \\ \hline
     Indexed &           0.0\% &             41.2\% &                  \$23,456 &            \$99,010 &        18.2\% \\ \hline
   Instadapp &           0.0\% &             39.7\% &                   \$1,843 &            \$38,878 &         9.6\% \\ \hline
     Inverse &           1.3\% &             46.7\% &                  \$41,775 &           \$120,220 &         8.3\% \\ \hline
PoolTogether &           6.1\% &             17.6\% &                  \$38,860 &           \$129,586 &        22.8\% \\ \hline
     Radicle &           1.5\% &             58.9\% &                   \$8,446 &           \$121,583 &        20.1\% \\ \hline
Rari Capital &           0.0\% &             24.7\% &                   \$7,673 &           \$244,402 &        31.7\% \\ \hline
        Silo &          29.4\% &             44.8\% &                   \$4,015 &            \$30,193 &        28.9\% \\ \hline
      Strike &            N/A &             62.2\% &                   \$4,693 &             \$29,583 &        1.9\% \\ \hline
     Uniswap &           6.1\% &             20.9\% &                 \$233,559 &         \$3,178,291 &         7.6\% \\ \hline

\end{tabular}
\caption{Characteristics of DAOs at blocks 16,530,000 (31 Jan 2023): the \emph{share of votes held by community delegates} (cf.\ Section \ref{sec:communityDelegates}), the \emph{participation rate of delegated voting power} (averaged over all proposals),
the \emph{cost of governance transactions} 
, the \emph{total cost of governance} (additionally including added overhead costs of transfers), and the \emph{proportion of useless transactions} (among delegation and vote transactions).}
\label{table:governance_overview}
\end{table}

\section{Conclusion}

The original promise of DAOs is to enable well accepted decisions by communities, allowing swift decision-taking both in times of great opportunity and difficult challenges. The proposals we observe and capture in our analysis range from suggesting protocol improvements, managing acquisitions and mergers, to handling the aftermath of hacks and economic downturn, sometimes even governing their own shutdown.

While open deliberation and voting is unquestionably good, we also observe a variety of alarming signs. By measuring low decentralization we find evidence that DAOs might be used as a marketing tool, or worse yet, as means to justify and veil decisions of a ruling dictatorship behind the facade of a community.

We hope that the shortcomings we lay bare can help inspire future DAO designs to also be more cost-effective and user-friendly. Finally, in the context of designing airdrop mechanisms, and in the wake of account abstraction\footnote{\url{https://eips.ethereum.org/EIPS/eip-4337}} allowing for governance incentivization, our analysis sheds light on resulting costs and decentralization in various scenarios and can thus help the creation of the next generation of DAOs.


\clearpage

\printbibliography

\newpage

\appendix

\section{Short Description of Analyzed DAOs}
\label{appendix:daos}

\paragraph{\href{https://uniswap.org/}{Uniswap}} is the market leading decentralized exchange on Ethereum. 
\paragraph{\href{https://compound.finance/}{Compound}} is a decentralized on-chain money market and lending platform. 
\paragraph{\href{https://ens.domains/}{ENS}} stands for Ethereum Name Service, a distributed service mapping human-readable addresses to wallet addresses for example.  
\paragraph{\href{https://www.gasdao.org/}{Gas DAO}} allows performing surveys among Ethereum users.  

\paragraph{\href{https://www.ampleforth.org/}{Ampleforth}} is a cryptocurrency with an algorithmically adjusted circulating supply. 

\paragraph{\href{https://gitcoin.co/}{Gitcoin}} is a platform designed to fund and govern open source projects. 

\paragraph{\href{https://fei.money/}{Fei}} is an algorithmic stablecoin. The protocol was governed by the Tribe DAO, but after acquiring Rari Capital and repaying Hack victims, it wound down its operations, and is now discontinued. The governing Tribe DAO is said to have pioneered both the first merger and the first wind-down of a protocol in the DeFi space. 

\paragraph{\href{https://hop.exchange/}{Hop}} is a protocol that allows transferring tokens between roll-ups. 

\paragraph{\href{https://strike.org/}{Strike}} is a lending and borrowing protocol. Intriguingly, all governance proposals have been created by the same account, and apart from the first proposals this same account is the only account that has ever voted (with exactly 131'000 tokens where the quorum of votes needed to pass a proposal consists of 130'000 votes).

\paragraph{\href{https://pooltogether.com/}{PoolTogether}} is a protocol that awards lottery prizes to participants.  

\paragraph{\href{https://www.rari.capital/}{Rari Capital}} is a DeFi lending and borrowing platform. It has been the target of two large scale attacks, the first draining 15 million USD, the second around 80 million USD. After being acquired by Fei, its governance token was also \$Tribe. 

\paragraph{\href{https://radicle.xyz/}{Radicle}} provides infrastructure for decentralized software collaboration. 

\paragraph{\href{https://indexed.finance/}{Indexed}} was a project that provided passive portfolio management strategies for the Ethereum ecosystem. After an exploit drained a large fraction of the locked assets, the governing DAO held multiple votes regarding lawyer payment and token refunds to users.

\paragraph{\href{https://idle.finance/}{Idle finance}} is a yield aggregator, offering different yield generating strategies to users. The governing DAO protocol was updated in January 2022.
Both the old and the new governance contract and incorporated in the dataset.

\paragraph{\href{https://instadapp.io/}{Instadapp}} aims a providing infrastructure to improve the DeFi user-experience through interfaces and simplified protocols. 

\paragraph{\href{https://www.usebraintrust.com/}{Braintrust}} is an online hiring marketplace for freelancing governed by the BTRST token.

\paragraph{\href{https://www.silo.finance/}{Silo}} is a lending protocol, that allows the borrowing of any asset with another. Its governance token is called SILO. 

\paragraph{\href{https://www.inverse.finance/}{Inverse}} is a protocol that generates yield on stablecoins and allows re-investment of the yield in a target token. We analyze the governing DAO before the smart contract was updated in October 2021.

\paragraph{\href{https://www.euler.finance/}{Euler}} is a lending protocol controlled by a DAO. Many proposals are happening purely off-chain. 

\paragraph{\href{https://cryptex.finance/}{Cryptex}} offers exposure to market capitalization of the crypto market at large. It is governed by holders of the CTX tokens. 

\paragraph{\href{https://docs.babylon.finance/getting-started/master}{Babylon}} was a community lead asset management protocol. After being affected by the Rari Fuse Hack mentioned above, the protocol shut down in November 2022. After some back and forth between different communities and DAOs, Babylon users were eventually repaid their lost funds. The Babylon team itself bought Tribe DAO tokens, in order to vote in favor of a proposal to refund money to hack victims.



\section{Computing the Cost of Governance}
\label{appendix:costs}

In order to estimate the governance costs, we extracted the gas required for the transaction and the gas price paid from the corresponding transaction receipts. 
Accurate measurements are complicated by the fact that the required gas for a transaction can vary widely, where delegation has taken place. More specifically, the gas requirement for a delegation varies depending on the address to which the delegation is made. In general, a delegation from delegate  $A$ to delegate $B$ with $ A \neq B $ requires more gas than a delegation from delegate $A$ to themselves. Line 9 in Listing \ref{lst:code} is the reason for this behavior. When a participant $A$ delegates to themselves, nothing changes regarding the delegates and therefore, the rest of the logic of the \textit{moveDelegates} function can be skipped. 

Another challenge we face is that the amount of used gas is only available for the entire transaction. However, a transaction can consist of several elements, e.g. a token claim and a subsequent delegation. In this case, if the gas consumed for the entire transaction were to be counted as governance costs, the estimate of the costs would be much higher than the effective costs. For such transactions we therefore only include the gas costs up to a fixed value, that we determine by computing the average amount of gas consumed for all isolated delegations - i.e. those transactions in which only a pure delegation was carried out. 

Finally, note that the costs of executing protocol changes following a successful proposal is not included in our computed price of governance, as these costs are incurred no matter the type of governance, be it on- or off-chain.

\begin{figure}
\begin{lstlisting}[language=Solidity,caption={A simplified depiction of two Solidity functions in the Uniswap smart contract.},label={lst:code}]
function transferTokens(address src, address dst, uint96 amount) internal {
        balances[src] = sub96(balances[src], amount);
        balances[dst] = add96(balances[dst], amount);
        emit Transfer(src, dst, amount);
        moveDelegates(delegates[src], delegates[dst], amount);
    }
    
 function moveDelegates(address srcRep, address dstRep, uint96 amount) internal {
        if (srcRep != dstRep && amount > 0) {
            if (srcRep != address(0)) {
              #Omitted
            }

            if (dstRep != address(0)) {
          		#Omitted
            }
        }
    }
\end{lstlisting}
\end{figure}

\section{Computing the Overhead Cost of Governance}
\label{appendix:overhead_costs}

In this section we outline the reason why a simple token transfer incurs higher costs if the token gives voting rights in a DAO.

In current DAOs there can be a maximum of one delegate at any given time for each address and all tokens held by that address. In other words, the voting power associated with the token of an address cannot be divided among different delegates. As a consequence, the transfer function from the ERC-20 standard \footnote{\url{https://eips.ethereum.org/EIPS/eip-20}} must be modified. Let $A,B,C,D$ be different addresses and $A$ delegating to $C$ and $B$ delegating to $D$. If $B$ transfers its tokens to $A$, then $A$ would have tokens whose voting power is delegated to $C$ and tokens whose voting power is delegated to $D$, which contradicts the constraint that there can be at most one delegate for each address. For this reason, during each transfer it must be checked whether a delegation is necessary and if so, this delegation must be carried out. Consequently, more gas is needed for the modified transfer than for the ERC-20 standard transfer. 

Thus, regardless of whether there is active participation in governance, the gas cost of a transfer increases. 

We have estimated these costs with the help of remix \footnote{\url{https://remix.ethereum.org}}. The implementation of the \textit{transferToken} function and the \textit{moveDelegates} function is very similar between different projects. First, we measured the cost of calling a \textit{transferToken} function as shown in Listing \ref{lst:code}. Then we measured how the cost changes when line 5 in Listing \ref{lst:code} is removed. We estimate the additional cost of the modified transfer function to be about 4500 gas. 

We multiplied the gas price at the time of a transfer by the estimated additional cost of 4500 gas. We then converted this value to USD using the same method as in the previous sections. 

\end{document}

%% file: solidity-highlighting.tex
\usepackage{listings, xcolor}

\definecolor{verylightgray}{rgb}{.97,.97,.97}

\lstdefinelanguage{Solidity}{
	keywords=[1]{anonymous, assembly, assert, balance, break, call, callcode, case, catch, class, constant, continue, constructor, contract, debugger, default, delegatecall, delete, do, else, emit, event, experimental, export, external, false, finally, for, function, gas, if, implements, import, in, indexed, instanceof, interface, internal, is, length, library, log0, log1, log2, log3, log4, memory, modifier, new, payable, pragma, private, protected, public, pure, push, require, return, returns, revert, selfdestruct, send, solidity, storage, struct, suicide, super, switch, then, this, throw, transfer, true, try, typeof, using, value, view, while, with, addmod, ecrecover, keccak256, mulmod, ripemd160, sha256, sha3}, 
	keywordstyle=[1]\color{blue}\bfseries,
	keywords=[2]{address, bool, byte, bytes, bytes1, bytes2, bytes3, bytes4, bytes5, bytes6, bytes7, bytes8, bytes9, bytes10, bytes11, bytes12, bytes13, bytes14, bytes15, bytes16, bytes17, bytes18, bytes19, bytes20, bytes21, bytes22, bytes23, bytes24, bytes25, bytes26, bytes27, bytes28, bytes29, bytes30, bytes31, bytes32, enum, int, int8, int16, int24, int32, int40, int48, int56, int64, int72, int80, int88, int96, int104, int112, int120, int128, int136, int144, int152, int160, int168, int176, int184, int192, int200, int208, int216, int224, int232, int240, int248, int256, mapping, string, uint, uint8, uint16, uint24, uint32, uint40, uint48, uint56, uint64, uint72, uint80, uint88, uint96, uint104, uint112, uint120, uint128, uint136, uint144, uint152, uint160, uint168, uint176, uint184, uint192, uint200, uint208, uint216, uint224, uint232, uint240, uint248, uint256, var, void, ether, finney, szabo, wei, days, hours, minutes, seconds, weeks, years},	
	keywordstyle=[2]\color{teal}\bfseries,
	keywords=[3]{block, blockhash, coinbase, difficulty, gaslimit, number, timestamp, msg, data, gas, sender, sig, value, now, tx, gasprice, origin},	
	keywordstyle=[3]\color{violet}\bfseries,
	identifierstyle=\color{black},
	sensitive=false,
	comment=[l]{//},
	morecomment=[s]{/*}{*/},
	commentstyle=\color{gray}\ttfamily,
	stringstyle=\color{red}\ttfamily,
	morestring=[b]',
	morestring=[b]"
}